\documentstyle[12pt,epsf,rotate]{article}
\catcode`\@=11
\def\subsection{\@startsection{subsection}{2}{\z@}{-3.25ex plus -1ex minus 
                -.2ex}{1.5ex plus .2ex}{\bf }}
\catcode`\@=12
\def\beq{\begin{equation}}      \def\eeq{\end{equation}}
\def\bea{\begin{eqnarray}}      \def\eea{\end{eqnarray}}
\def\bq{\begin{quote}}          \def\eq{\end{quote}}
   
\renewcommand{\(}{\left(} \renewcommand{\)}{\right)} 
\renewcommand{\[}{\left[} \renewcommand{\]}{\right]} 
\def\tr{\mathop{\rm tr}} 
 
\def\Re{\mathop{\rm Re}}
\def\Im{\mathop{\rm Im}}

\def\lta {\mathrel{\vcenter {\hbox{$<$}\nointerlineskip\hbox{$\sim$}}}}
\def\gta {\mathrel{\vcenter {\hbox{$>$}\nointerlineskip\hbox{$\sim$}}}}

\def\gtap{\raisebox{-.4ex}{\rlap{$\sim$}} \raisebox{.4ex}{$>$}}
\def\slash#1{#1\!\!\!/\!\,\,}
\renewcommand{\bar}[1]{\overline{#1}}

\def\phi{\varphi}

\def\sixtpi{\frac{1}{16\pi^{2}}}
\def\SIXTPI{16\pi^{2}}
\def\lnlambda{\ln \left( \frac{\Lambda^{2}}{\mu^{2}} \right)}
\def\vr2{v_{R}^{2}}
\def\kap2{\kappa^{2}}
\def\kapp2{\kappa'^{2}}
\def\hvr{\hat{v}_{R}}
\def\hvr2{\hat{v}_{R}^{2}}
\def\hkap2{\hat{\kappa}^{2}}
\def\hkapp2{\hat{\kappa'}^{2}}

            \def\dash{\hbox{---}}
\def\tilde{\widetilde}
\def\SMo{Standard Model}
\def\SM{\SMo\ } \def\SMp{\SMo.\ } \def\SMk{\SMo,\ }
\def\EW{electro-weak } 
\def\GBo{Goldstone boson} \def\GBsk{\GBo s,\ } \def\GB{\GBo\ } 
 \def\GBs{\GBo s\ }  \def\GBsp{\GBo s.\ }
\def\RG{renormalization group } 
\def\DSBo{dynamical symmetry breaking} \def\DSB{\DSBo\ } 
 \def\DSBk{\DSBo,\ }
\def\NJLo{Nambu--Jona-Lasinio} \def\NJL{\NJLo\ } 
\def\LRo{left-right} \def\LR{\LRo\ } 
\def\LRS{\LRo-symmetric\ }
\def\SMLo{\SM limit} 
\def\SML{\SMLo\ } \def\SMLk{\SMLo,\ } \def\SMLp{\SMLo.\ }
\begin{document}

%
\begin{flushright}
TUM--HEP--293/97\\
LBNL--41081 \\
\end{flushright}
\ \vskip 1.cm 
%
%

\begin{center}

{\Large\bf Dynamical Electro-Weak Symmetry Breaking}

{\Large\bf with a Standard Model Limit~$^\dagger$}

\vspace*{5mm}

{\large\bf Manfred Lindner$^*$ and Erhard Schnapka$^{**~\ddagger}$}

\vspace*{5mm}

\renewcommand{\thefootnote}{\fnsymbol{footnote}}

\footnotetext[2]{To appear in ``Heavy Flavours II'', 
eds. A.J. Buras and M. Lindner, Advanced Series on Directions in 
High Energy Physics, World Scientific Publishing Co., Singapore}

\footnotetext[3]{Supported by the 
             BASF AG and the Studienstiftung des deutschen Volkes.}

\renewcommand{\thefootnote}{\alph{footnote}}

{\sl
$^*$ 
Physik Department, Technische Universit\"at M\"unchen\\
James-Franck Str., D-85748 Garching b. M\"unchen, Germany\\
Email: lindner@physik.tu-muenchen.de\\
\ \\
$^{**}$ 
Theoretical Physics Group\\
Ernest Orlando Lawrence Berkeley National Laboratory\\
University of California, Berkeley, California 94720, USA\\
Email: schnapka@lbl.gov\\ 
}

\vspace*{2.cm}

\begin{minipage}[t]{12.5cm}
We argue that a \SM decoupling limit is generically the necessary 
ingredient which makes scenarios of \EW symmetry breaking viable. 
This applies especially also to models of dynamical \EW symmetry 
breaking. Additional requirements are only that the mass predictions 
of a given model (e.g. predictions or theoretical limits on the 
Higgs or top mass) are consistent with existing data. We discuss 
the necessary ingredients for \DSB and present a dynamically broken 
\LRS model as an example. The model exhibits such a decoupling limit, 
is phenomenologically viable and leads to interesting mass 
predictions and relations which are further examined.
\end{minipage}

\end{center}

\vspace{2.cm}


\newpage

\section{Introduction}

In recent years the \SM has been able to describe the 
experimental data with impressive accuracy~\cite{SMreview}, 
including various different reactions and precision tests of 
radiative corrections. Despite the successes of the \SMk the 
current experiments mainly test the gauge structure. For the 
Higgs sector we know essentially only that the gauge symmetry 
is broken by some suitable vacuum expectation value. In the 
\SM the elementary Higgs sector leads to the famous hierarchy 
problem~\cite{HP}, due to the quadratic divergences associated 
with fundamental scalar particles. A solution to this problem 
probably involves the embedding of the \SM in new physics at the 
TeV-scale, and there are two main approaches: Either the quadratic 
corrections are cancelled due to restored supersymmetry (SUSY) 
above the SUSY breaking scale $\Delta\simeq TeV$, or some strongly 
interacting dynamical \EW symmetry breaking scenario provides 
form-factors (i.e. un-binding) and eliminates elementary Higgs 
bosons in the underlying theory.

Both routes have attractive features, but what can we learn 
from the current experimental situation? Within the framework of 
SUSY-GUTs one can obtain a remarkably accurate prediction of the weak 
mixing angle from gauge coupling unification. However, although being a 
nice feature, this is hardly direct evidence for supersymmetry. Apart 
from that, the so-called Minimal Supersymmetric \SM (MSSM) is unchallenged 
by current experimental data. This is however largely because in the MSSM 
all non-\SM effects can decouple if the soft supersymmetry breaking mass 
parameters are raised in the TeV-region and flavour changing effects are 
tuned to be small. In other words, the MSSM can hide sufficiently behind 
the \SMk at least as long as the experimental situation allows the Higgs 
mass to be below the MSSM-upper bound.

The situation may appear different for dynamical \EW symmetry breaking 
with a strongly interacting sector in the TeV-region. In {\em naive} 
Technicolor models, based on a rescaled version of QCD, one finds severe 
constraints from the existing data, and most of the original models are 
now ruled out by experiment. Note, however, that this is not a generic 
problem of \DSBk but essentially the failure of {\em any} model where 
the effective Lagrangian does not have a \SMo-limit. Both perturbative 
or non-perturbative scenarios of physics beyond the \SM with a decoupling 
limit are, on the other hand, phenomenologically acceptable, unless 
Higgs or other mass predictions disagree with experimental constraints. 
The question is thus if models of \DSB with a \SML can be built.

There are genuine {\em technical} differences between low energy 
supersymmetry and models of dynamical \EW symmetry breaking which 
should not be used as an argument for or against either direction. 
While the MSSM is mostly perturbative and, therefore, its phenomenology 
relatively easy to analyse, \DSB generically 
arises non-perturbatively. Therefore many quantities in such models are 
harder to calculate and predictions often very rough~\footnote{For 
example, in a recently proposed Technicolor-like model~\cite{AWT} based 
on an approximate infrared fixed point and near criticality, the 
additional effects of ordinary QCD can lead to \EW symmetry breaking. 
However, lattice simulations or better analytical understanding of 
non-perturbative effects will be needed to confirm the phenomenological 
viability of such a framework.}. There is also no or little guidance from 
a greater picture compared to a supersymmetric framework. 

Important are the {\em conceptual} differences between the approaches 
and ultimately, of course, future experiments will have to decide which 
scenario is realized in nature. Low energy supersymmetry may radiatively 
induce \EW symmetry breaking and leads to a rich phenomenology which can 
be studied perturbatively. At the same time, the hierarchy problem 
can be solved, provided that one finds a compelling (dynamical) model of 
supersymmetry breaking that leads to the desired soft-breaking terms 
in the TeV-region. 
As an alternative to models of supergravity-mediated 
SUSY-breaking~\cite{SUGRA}, a variety of models have been proposed 
where SUSY breaking is mediated at low energies by standard gauge 
interactions~\cite{gaugemed}. Building such models has been helped 
by a significant improvement in understanding the non-perturbative dynamics 
of N=1 supersymmetric gauge theories~\cite{seib-intr}.
On the other hand, ``immediate'' dynamical \EW symmetry breaking 
is attractive from a simplistic point of view. Dynamical 
symmetry breaking is a natural effect in strongly interacting theories 
when an asymmetric ground state lowers the total energy of the system. 
Well known examples in other fields of physics are ferro magnetism and 
superconductivity, where in the latter case one finds a dynamical Higgs 
mechanism. Unfortunately, far less is know about the non-perturbative 
behaviour of {\em non}-supersymmetric field theories, compared to the 
supersymmetric case. Nevertheless it is worth to consider scenarios where 
the \EW gauge symmetry is dynamically broken by strong dynamics in the 
TeV-region. There might even be a sequence of layers of new physics, 
which perhaps also involves supersymmetry at scales well beyond the \EW 
scale. 

In this article we would like to exemplify, that the {\em current} 
experimental situation essentially favours {\em any} framework which 
can ``hide'' behind the \SMp We will argue that it is possible to 
systematically build models of \DSB with such 
a decoupling limit, which by construction may be consistent with the 
stringent constraints from the data. As an example, such an intrinsic 
decoupling limit can be found in models of top-condensation, where 
the \SM Higgs boson is replaced by a scalar $\bar{t}t$-bound-state, 
acting as the unitarity partner for the \GBsp Later in this article 
we will present a phenomenologically viable dynamically broken \LRS 
model based on this approach.

\section{The Standard Model Limit}

By now there exists direct evidence for all fermions and gauge bosons 
of the \SM and we know that the \EW gauge group is broken. There are 
experimental lower limits and theoretical (indirect and model dependent) 
upper limits for the Higgs mass, but there is no direct evidence for 
the existence of a Higgs particle. Existing data provides, however, 
additionally numerous restrictions for modifications of the Higgs sector. 
Examples are rare decays, FCNC effects, contributions to $R_b$ and other 
indirect effects via radiative corrections. Among those the so-called 
oblique corrections, commonly parametrized by the S-, T- and U- parameters, 
are particularly important. Thus, while a model of new physics attempting 
to explain \EW symmetry breaking need not feature an elementary Higgs 
boson, it certainly must be in accordance with all the above mentioned 
indirect constraints which are consistent with the \SMp

We think of a model of new physics as having a ``\SM limit'',
if the parameters of the model can be chosen such that (within 
current experimental errors) the model becomes in a certain limit
indistinguishable from the \SMk {e.g.}, all effects from 
additional particles and interactions can be decoupled. If this is 
possible, it will allow a certain range of those parameters which are so 
far only poorly determined within the \SMk such as the mass 
of the \SM Higgs boson, certain mixing angles, details of CP violation 
or neutrino masses. Thus a \SML - if it exists - does not necessarily 
lead to every possible choice of \SM parameters, and examples for `relic' 
mass restrictions ({e.g.}, for the Higgs mass) will be seen below.

The minimal supersymmetric extension of the \SM (MSSM)
exhibits such a \SMLp To be phenomenologically 
viable, one must ensure sufficient separate conservation of the 
three lepton numbers, the suppression of flavour changing neutral 
currents (FCNC) as well as the smallness of the neutron and 
electron electric dipole moments. As emphasized in a recent 
review~\cite{haber97}, this confines the MSSM to special, ``exceptional'' 
points in parameter space, and there are several attempts to explain 
this situation~\footnote{In the MSSM with minimal particle content and 
R-parity conservation assumed, the number of parameters of the model
equals 124~\cite{haber97}. One possibility to largely reduce this set of 
parameters, and to naturally  solve the SUSY-flavour problem in the process,
opens up in the context of gauge mediated supersymmetry breaking.
The flavour blindness of gauge interactions leads to flavour blind 
soft breaking squark and slepton mass terms, which ensure the desired FCNC 
suppression.}. Once within this parameter regime, the virtual effects of 
supersymmetric particles can be decoupled from \EW observables at or 
below the $Z$-mass~\cite{mssmdecoupl}, when the masses of all 
supersymmetric particles are raised above $\sim 200$~GeV. Especially, 
all the non-minimal Higgs bosons of the MSSM ($H^0$, $H^\pm$ and $A^0$) 
can be decoupled, when the mass of the CP-odd Higgs boson is raised 
($m_{A^0} \gg m_{Z^0}$). In this limit the CP-even Higgs boson $h^0$ 
remains light, with \SM equivalent couplings to all \SM particles. This 
is because supersymmetry relates the quartic Higgs coupling (and mass) to 
gauge couplings, and after taking radiative corrections into account the 
CP-even Higgs bosons is always lighter than $\sim 125$~GeV~\cite{mssmhiggs}.  
Thus, due to this \SMLk the MSSM can be consistent with \EW precision data, 
provided the additional ``relic'' mass restriction is met. 
I.e., the global fit for the \SM Higgs mass must be consistent
with the upper bound on $h^0$, $m_{h^0} \lta 125$~GeV. 

Let us now see how the \SML can be realized in different scenarios with 
strongly interacting physics in the TeV-region with dynamical \EW symmetry 
breaking. Besides the issue of flavour changing neutral currents, it is 
very important to consider the so-called ``oblique radiative corrections'', 
which can be parametrized in the precision variables $S$, $T$, and $U$. 
These corrections are defined from the vacuum polarization amplitudes of 
$\gamma$, $W$, and $Z$ and the $\gamma Z$ mixing which have the form
\bea
\Pi_{\gamma\gamma} & = & e^2 \Pi_{QQ} ~,\\
\Pi_{WW} & = & \frac{e^2}{s^2} \Pi_{11} ~,\\
\Pi_{ZZ} &=& \frac{e^2}{c^2s^2} (\Pi_{33} - 2s^2 \Pi_{3Q}+ s^4\Pi_{QQ})~, \\
\Pi_{Z\gamma} &=& \frac{e^2}{cs} (\Pi_{3Q} - s^2 \Pi_{QQ})~.
\label{oblique}
\eea
Here the relation $J_Z  = J_3 - s^2 J_Q$ has been used. $J_Q$ is the 
electro-magnetic current, $s^2 = \sin^2\theta_w$, $c^2 = \cos^2\theta_w$, 
and the weak coupling constants have been expressed in terms of $e$, $s^2$ 
and $c^2$. The indices $i,j$ of $\Pi_{ij}$ on the right hand side indicate 
the relevant currents. The \SM contributions to the vacuum polarizations can 
be separated such that the remaining contributions of new physics are then 
functions of $q^2/M^2$, where $M^2$ is the mass of new, heavy particles. 
If $M^2$ is large enough one can expand in powers of $q^2$. Using QED 
Ward-Identities for $q^2=0$ the expansion leads at order $q^2$ to six 
coefficients:
\bea
\Pi_{QQ} & = & 
\ \ \ \ \ \ \ \ \ \ \ \ q^2 \Pi^{'}_{QQ}(0) + .... \\
\Pi_{11} & = & 
\Pi_{11}(0) +  q^2 \Pi^{'}_{11}(0)\ + .... \\
\Pi_{3Q} & = & 
\ \ \ \ \ \ \ \ \ \ \ \ q^2 \Pi^{'}_{3Q}(0)\ + .... \\
\Pi_{33} & = & 
\Pi_{33}(0) +  q^2 \Pi^{'}_{33}(0)\ + ....
\label{PIij}
\eea
When the three most precise measured observables, $\alpha_{em}$, $G_F$ and 
$m_Z$, are used as input there remain three independent variables which 
parametrize effects of new physics. These three variables can be defined 
as~\cite{STUref}
\bea
S&=& 16\pi\[\Pi^{'}_{33}(0) -\Pi^{'}_{3Q}(0)\] ~,\\
T&=& \frac{4\pi}{s^2 c^2 m_Z^2}\[\Pi_{11}(0) -\Pi_{33}(0)\]~, \\
U&=& 16\pi\[\Pi^{'}_{33}(0) -\Pi^{'}_{11}(0) \] ~.
\label{STU}
\eea
The last variable, $U$, is so small that it is currently irrelevant and 
will not be considered further. The variable $T$, measuring custodial SU(2) 
violating effects, and $S$, which is related to axial $\mbox{SU(2)}$ and 
which is sensitive to the unitarity partner of \GBsk will both be discussed 
in more detail below. 

It is instructive to consider the effects of naive Technicolor scenarios 
on oblique radiative corrections. The original models of this type, as well 
as more sophisticated versions like extended Technicolor (ETC) or 
``walking'' Technicolor, have largely been ruled out in the 
past~\cite{TCruleout}, and in the following we describe the basic 
difficulties. In the simplest version of these models of dynamical \EW 
symmetry breaking, the \GB decay constant of QCD is scaled up such that 
the correct $W$ mass arises. This leads (up to $N_{TC} \neq 3 = N_c$ 
corrections) to a more or less fixed spectrum of QCD-like bound states in 
the TeV-region. Consequently the Techni-pions, the \GBs giving mass to 
$W$ and $Z$ in a dynamical Higgs mechanism, do not have a Higgs-like, 
{\em i.e., scalar}, unitarity partner~\footnote{The so-called sigma 
particle would -- if it exists -- be broad and too heavy.}. Instead, the 
role of unitarity partners is played predominantly by low lying composite 
vector resonances, the Techni-rhos. Due to the QCD-like dynamics it is 
therefore not possible to obtain the \SM spectrum, where a suitable 
composite Higgs-like scalar mimics the \SM Higgs boson, and the remaining 
spectrum is decoupled. However, the main phenomenological problem of 
naive Technicolor is not per se the absence of the scalar partner of the 
\GBsk but the low mass of the Techni-rho resonances. We will see in a 
moment that low lying (composite or fundamental) vector-like states are 
in general a problem for the experimentally small precision variable $S$, 
even if there is a (composite or fundamental) Higgs particle. If a 
Higgs-like scalar is absent, like in Technicolor, then there exist of 
course upper bounds for the masses of vector states due to the unitarity 
of \GB scattering amplitudes. Such upper bounds for the masses of vector 
states do not exist in models where a Higgs particle exists in addition 
to extra vector states. An example is given by models which contain a 
$Z^\prime$, which can in principle become arbitrarily heavy.

The phenomenological problem with light vector states is that they 
can mix with the $W$ and $Z$-bosons, which is severely constrained 
by precision \EW data, i.e. the smallness of the $S$ parameter. 
This explains why in Technicolor the problem becomes even more severe as 
the number of Techni-colors $N_{TC}$ is increased, since for large 
$N_{TC}$ the ratio of Techni-rho mass and \GB decay constant 
becomes smaller, and the mixing with $W$ and $Z$ is 
increased. This effect can be seen best by expressing $S$ with 
the help of dispersion relations~\cite{STUref} as
\beq
S=\frac{1}{3\pi}\int^\infty_0\frac{ds}{s}
\[ R_V(s) - R_A(s) - H(s)\]~,
\label{Sformula}
\eeq
where $R_V$ and $R_A$ measure the contribution of vector and
axial-vector states, respectively. $R_V$ and $R_A$ 
are defined as the ratios of the cross sections of a photon 
which couples to the isospin current $J^{\mu 3}$ divided by 
the Compton process, in analogy to the famous $R$ ratio in QCD.
The definition of $S$ depends on a reference \SM Higgs mass 
value which was often chosen to be $300$~GeV. The function 
$H(s)$ allows one to remove the \SM Higgs contribution, e.g., 
in Technicolor
by sending the Higgs mass into the continuum of bound states 
(i.e., $m_H\simeq 1$~TeV). The function $H$ can be written as
\beq
H(s) = \frac{1}{4}\(1-\frac{m_H^2}{s}\)^3\theta(s-m_H^2) 
\label{Hs}
\eeq
which leads essentially to a small logarithmic dependence on 
the Higgs mass value.

In Technicolor the quantities $R_V$ and $R_A$ measure the sum of 
charges involved in $I=1$ vector resonances. One can estimate
$S$ by parametrizing (similar to QCD) the resonance and the 
continuum contributions to $R_V$ and $R_A$ by delta and theta 
functions, respectively. The leading contributions come from the 
vector meson resonances (vector meson dominance), which leads to
\beq
R_V(s) = 12 \pi^2 F^2_{\rho T} ~\delta(s-m^2_{\rho T})~, \quad
R_A(s) = 12 \pi^2 F^2_{a_1 T}  ~\delta(s-m^2_{a_1 T})~,
\label{RVRA}
\eeq
where $m_{\rho T}$ and $m_{a_1 T}$ are the Techni-rho and Techni-$a_1$ 
masses. With the help of the Weinberg sum rules~\cite{Weinberg} it is 
possible to express $F^2_{\rho T}$
and $F^2_{a_1 T}$ as
\beq
F^2_{\rho T}= \frac{m^2_{a_1 T}F_\pi^2}{m^2_{a_1 T}-m^2_{\rho T}}~,
\quad
F^2_{a_1 T} = \frac{m^2_{\rho T}F_\pi^2}{m^2_{a_1 T}-m^2_{\rho T}}~,
\eeq
where $F_\pi=250$~GeV. From this one obtains from $R_V$ and 
$R_A$ the following contribution to $S$:
\beq
S=4\pi\( 1+\frac{m^2_{\rho T}}{m^2_{a_1 T}}\)\frac{F_\pi^2}{m^2_{\rho T}}~.
\label{Sfromdelta}
\eeq
Using large $N_c$ rescaling relations between Technicolor and QCD one 
finds~\cite{largeNc}
\beq
\frac{m^2_{\rho T}}{m^2_{a_1 T}} = \frac{m^2_{\rho}}{m^2_{a_1}}~
\quad
\frac{F_\pi^2}{m^2_{\rho T}} = 
N_D \frac{N_{TC}}{3}\frac{f_\pi^2}{m^2_{\rho}}~.
\eeq
With $f_\pi=93$~MeV, $m_\rho=770$~MeV and $m_{a_1}=1260$~MeV this results 
in $S\simeq 0.25 N_D N_{TC}/3$, where $N_{TC}$ is the number of 
Techni-colors and $N_D$ is the number of weak doublets. The continuum 
contributions to $R_V$ and $R_A$ can be parametrized by theta-functions 
and lead to logarithmic mass dependencies, while the resonances lead to 
$1/m^2_{\rho T}$ contributions. These $1/m^2_{\rho T}$ contributions are 
usually dominant unless logarithmic contributions proportional to 
$N_{TC}^2$ become for $N_{TC} > 8$ equally or even more important.

If one evaluates $R_V$ and $R_A$ using a more detailed analysis
of QCD data and large $N_c$ rescaling one finds~\cite{STUref}
\beq
S =  0.3 N_D ~\frac{N_{TC}}{3} ~,
\label{STC}
\eeq
which must be confronted with the value of $S$ extracted from global fits 
to existing data. For such fits it is necessary to specify the top quark 
mass and the Higgs mass of the \SMk and typically the values 
$m_t = 175$~GeV and $m_H = 300$~GeV are assumed~\cite{Sexp}. For 
comparison with Technicolor models the reference Higgs mass is sometimes 
shifted to $1$~TeV, which leads to 
\beq
S = -0.26 \pm 0.16~.
\label{Svalue}
\eeq
The negative sign expresses the fact that the preferred Higgs mass of 
the \SM is much smaller than 1~TeV, and today the best fit is 
obtained~\cite{bestH} for $m_H \approx 121$~GeV. Comparing 
equation~(\ref{Svalue}) with the prediction of naive Technicolor, 
equation~(\ref{STC}), one finds the smallest disagreement at the 
$3~\sigma$ level for $N_D =1$ and $N_{TC} =3$. More Technicolor 
doublets and/or a larger Technicolor group lead to even larger
deviations from the data.

The above expression for $S$, equation~(\ref{Sformula}), applies in 
principle for any theory with composite or fundamental vectors and/or 
axial-vector states which couple to $J^{\mu 3}$. For example, one could 
study $Z^\prime$ models in this way and find a $1/m^2_{Z'}$ dependence 
from the mixing of the $Z^\prime$ with $Z$. This leads to unacceptable 
contributions to $S$ if the extra vector state becomes light. However, 
these contributions are not as severe as in Technicolor since the 
$Z^\prime$ couplings are usually chosen to be perturbative (i.e. small) 
and the color factor $N_{TC}$ is absent. 

Another severe problem of many Technicolor scenarios stems from
additional extra Techni-fermion doublets. The problem shows up in the 
precision variable $T$ defined above and emerges in a more general 
context for new fermionic doublets, i.e., it is not only a genuine 
Technicolor problem. Since extra fermions are not observed such a
$\mbox{SU(2)}_L$ doublet is required to be massive with masses
well above one hundred GeV. When the massive propagators for 
extra doublets of fermions are written as 
\beq
S_j = \frac{i}{\slash{p}-\Sigma_j(p^2)} ~,
\label{Dprop} 
\eeq
where $j=U,D$ for a new doublet and $j=t,b$ for the top and bottom 
quark, then the contribution to $T$ is given by~\cite{BlLi}
\beq
T=\frac{-N_c}{16\pi^2\alpha_{em}v^2}
~\int\limits_0^\infty dk^2~
\frac{k^4(\Sigma_U^2-\Sigma_D^2)^2}
{(k^2-\Sigma_U^2)^2(k^2-\Sigma_D^2)^2}~.
\label{Tdoublet}
\eeq
Here $N_c$ is the number of colors or Techni-colors and
$v\simeq 246$~GeV. If we insert $\Sigma_t=m_t$ and $\Sigma_b=0$ 
into equation~(\ref{Tdoublet}) one finds the well
known, leading \SM top mass contribution to the $T$-parameter,
\beq
T_{SM}
= \frac{N_c}{16\pi^2\alpha_{em}}~\frac{m_t^2}{v^2}
= \frac{N_c}{16\pi\sin^2\theta_W\cos^2\theta_W}
  ~\frac{m_t^2}{M_Z^2}~.
\label{SMT}
\eeq
Equation~(\ref{Tdoublet}) is also valid for new heavy doublets. 
In Technicolor, for example, ordinary quark and lepton masses (like 
the top mass) must be generated by so-called Extended 
Technicolor~\cite{ETC} interactions between quarks and 
Techni-quarks. The coupled system of gap equations leads, 
in a first approximation~\cite{AppWi}, to the relation 
\beq
\Sigma_U-\Sigma_D = \Sigma_t ~,
\label{sigmaETC}
\eeq 
where $\Sigma_U$ and $\Sigma_D$ are the mass functions of the 
Technicolor doublet. Assuming that equation~(\ref{sigmaETC}) is valid 
and approximating $\Sigma_i$ as $\Sigma_i=m_i\,\Theta(\Lambda^2-p^2)$, 
equation~(\ref{Tdoublet}) yields corrections to the \SM value of $T$:
\bea
T_{total} &=& T_{SM}+T \\
          &=& \frac{N_cm_t^2}{16\pi^2\alpha_{em}v^2}
\left(1 + \frac{4}{9} N_{TC} + \frac{m_t^2}{\Lambda^2}
+\frac{4}{3} N_{TC}\frac{m_U^2}{\Lambda^2}
+{\cal O}(m^4/\Lambda^4) \right)~ \nonumber 
\label{TTC}
\eea
For $\Lambda\rightarrow\infty$ this expression becomes the result 
usually quoted in the literature~\cite{AppWi}. For finite 
$\Lambda$ we can read off the $m_t^2/\Lambda^2$ correction to 
equation~(\ref{TTC}), and in addition a term proportional to 
$m_U^2/\Lambda^2$. For sufficiently large $\Lambda$ these 
$1/\Lambda^2$ terms are small and can be omitted. 
In this limit the value of $T$ is given by $4/9\cdot N_{TC}\cdot T_{SM}$, 
which is excluded by phenomenology due to the excellent 
agreement of the experimental value of $T$ with its \SM value. 
One might argue that equation~(\ref{sigmaETC}) is very model dependent
and may be generalized by a more complex relation, but it 
is not easy to obtain the top-bottom mass splitting 
without sizable corrections to $T$. 

As mentioned earlier, the problems of extra doublets in Technicolor 
have a more general scope. Any theory with extra doublets and a 
similar mechanism to explain the top-bottom mass difference faces 
similar problems. If the mechanism which explains the top-bottom mass 
splitting induces a $U-D$ splitting proportional to $K\cdot m_t$, where 
$K$ is typically expected to be of order unity, then this will lead to 
$T= 4/9\cdot K^2 N_c ~T_{SM}$. Unless $K$ is very tiny, this leads again
to amounts of custodial SU(2) violation (measured by $T$) too large to 
be reconciled with precision measurements. 

This discussion shows that two ingredients are disfavoured when 
constructing models of \DSB with a \SMLo : 
Low lying vector states lead to undesired contributions to the 
S-parameter, and additional fermionic SU(2)-doublets 
cause excessive violation of custodial SU(2) symmetry, unless
they are degenerate in mass.

At the end of this section we would like to emphasize, that neither
the \SML properties of the MSSM nor the problems with naive Technicolor
stem generically from a supersymmetric or dynamical mechanism of
\EW symmetry breaking. The phenomenological viability or
failure is to a large extent simply the presence or absence of the
mentioned \SMLk due to the lack of deviations of the experimental
precision data from the \SMp  In other words, this can be
seen as the success of models which have a limit, where the low energy
Lagrangian becomes effectively the \SM with a single \SMo-like Higgs
boson. In this respect low lying vector particles and extra fermionic
doublets, as they appear in naive Technicolor-models, are disfavoured.

Note that it is irrelevant for this discussion whether the resonances 
(vector or scalar) are elementary or composite. Therefore it appears 
promising to develop models of \DSB with purely 
scalar unitarity partners for the \GBsp With an effective or composite 
Higgs state, instead of Techni-rho like vector resonances, such models 
can include a \SML in the sense described earlier, and thus generically 
have a better chance to be phenomenologically viable. 
Future precision measurements may ultimately confirm specific deviations
between the \SM and the data (e.g. in quantities like $R_b$, etc.). 
These deviations might then be explained more easily 
by departing from the \SMLk i.e., by lowering some of the 
additional states, which are expected close to the continuum. 
Alternatively, of course, they might rule out a particular model. 
In any case, the underlying dynamics of such a scenario will likely 
lead to restrictions in the mass of whatever assumes the role of the Higgs 
boson in the \SMp In the MSSM this is the CP-even scalar $h^0$ which mass
is bounded from above ($ \le 125$~GeV) by the underlying dynamics, in 
a dynamical scenario considered here it will be a composite Higgs boson. 

\section{Guidelines for Model Building} 

The attempt to construct a viable model of dynamical \EW symmetry 
breaking may be viewed as a bottom-up approach, guided by the current 
phenomenological situation. As we saw in the previous section, the data 
so far favours any model with a \SMLk and we will identify the ingredients 
of symmetry breaking that {\em naturally} lead to such a limit. In this 
context it is important to carefully distinguish between the Higgs mechanism
(i.e. the \GBs ``eaten'' by gauge bosons, serving as their longitudinal 
degrees of freedom) and a physical Higgs particle. While the scalar \GBs are 
simply a consequence of broken global symmetries, irrespective of the 
nature of symmetry breaking (perturbatively or tree level), the existence 
or non-existence of a Higgs particle is a feature of the particular field 
theory considered. 

The Higgs mechanism for a specific gauge symmetry breaking pattern requires 
only an operator $\hat O$ with the following properties:
\begin{itemize}
\item 
$\langle \hat O\rangle \neq 0$, i.e. a ``condensate''
\item
Lorentz invariance of the vacuum requires that 
$\hat O$ must be scalar 
\item 
$\hat O$ must transform non-trivially under the gauge group to be broken, 
and as a singlet under the desired unbroken subgroups. 
\end{itemize}
Note that $\hat O$ does not have to be a fundamental scalar Higgs field. 
For the case of the \SM gauge group 
$\mbox{SU(3)}_c\times \mbox{SU(2)}_L\times \mbox{U(1)}_Y$ the operator
$\hat O$ should be a doublet of $\mbox{SU(2)}_L$ with $Y=1$ for 
the hypercharge~\footnote{$\hat O$ could in principle have 
different quantum numbers, but phenomenological evidence
like the smallness of $\Delta\rho$ lead to strong constraints.}.
The condensate $\langle \hat O\rangle \neq 0$ then breaks 
the global $\mbox{SU(2)}_L\times \mbox{U(1)}_Y$ symmetry, which implies
the existence of \GBs which can be eaten by W's and Z's. 
In unitary gauge the operator can be expanded as
\beq
\hat O = \left(\langle \hat O\rangle + \delta \hat O\right) 
         e^{i\phi_a T_a}\, ,
\eeq
and if a $|D\hat O|^2$ term is present in the effective Lagrangian 
with $\langle \hat O\rangle \neq 0$, the \GBs $\phi_a$ are absorbed 
by the corresponding gauge bosons. The Higgs mechanism requires
only the existence of a suitable condensate $\langle \hat O\rangle \neq 0$
and operates on the basis of symmetries, it does not depend on either
the fundamental or non-fundamental nature of $\hat O$ and/or 
the presence of $\delta \hat O$. 

The interactions of the \GBs can thus be understood in terms of the 
involved symmetries and the corresponding Ward-Identities, and the 
details of the interaction responsible for symmetry breaking become 
almost irrelevant. This is well established in QCD, where even a \NJL 
description of chiral symmetry breaking (which has almost nothing 
to do with QCD, but can be arranged to break the chiral symmetries 
correctly) leads to a remarkable good description of pion interactions.

On the other hand, $\delta \hat O$ or other non-scalar excitations 
of the vacuum are not related to the symmetries of the theory and 
may or may not include a fundamental or composite physical scalar 
Higgs particle $H$. If $\hat O$ is fundamental and $\delta \hat O$ 
is omitted completely, one arrives at the non-linear sigma model. 
This is fine with respect to symmetry breaking, but renormalizability 
is lost, which is an essential feature of a fundamental scalar theory.  
Unless new physics is very close, \GB scattering amplitudes are 
unbounded and unitarity is violated. One way out is to postulate a 
fundamental scalar Higgs field $H$ which can be grouped together with 
the \GBs in a $\mbox{SU(2)}_L$ doublet $\Phi$, and where $H$ can act 
as ``unitarity partner'' for the \GBsp  Unless new physics (i.e. 
some extra scale) is close renormalizability requires in addition 
that the Higgs potential $V(H)$ has only a mass term and 
$\lambda \Phi^4$ interactions. 

Note what happens if $\delta \hat O$ corresponds to a composite operator. 
In this case the \GBs are composite as well, and the Higgs mechanism 
will work as before. However, the remaining spectrum of the theory will 
typically be rich with an effective interaction Lagrangian which need 
not be renormalizable~\footnote{An example is again chiral symmetry 
breaking in QCD in the limit where pions are massless. Only the 
underlying theory -- here QCD -- should be based on fundamental fields 
and be renormalizable.}. Whatever the spectrum of the theory is, 
the effective Lagrangian must contain some resonances which act as 
unitarity partner for the \GBsp The simplest scenarios would be either 
a composite Higgs like state (i.e. an effective \SMo) or suitable vector 
resonances with the remaining spectrum located close to the continuum, 
i.e., at some high scale $\Lambda$. The details of the spectrum remain 
a dynamical issue, depending on the interactions of the underlying model. 
In QCD-like theories, for example, rho-like vector resonances emerge 
in the effective Lagrangian while a physical Higgs particle is absent. 
One can, however, choose interactions which lead to a composite Higgs $H$ 
instead.  Note that the existence of a composite scalar $H$ allows a 
richer interaction potential and other resonances in addition. Thus a 
composite Higgs does not automatically lead to an effective \SMp  

\noindent
We can now discuss how the \SML can be reached systematically in \DSB models:
\begin{itemize}
\item
At first a suitable symmetry breaking pattern must be considered 
in order to arrive at the correct spectrum of \GBsp Care should be 
taken to avoid problems with so-called pseudo \GBsk since, being 
light, they lead easily to phenomenological problems. 
\item
The existing data for the precision variable $S$ favours a spectrum where 
a Higgs-like scalar plays the role of the unitarity partner of the \GBsk
as explained in section 2. Vector-like states with SU(2)-quantum numbers 
are disfavoured, and thus one is lead to consider scenarios which differ
significantly from QCD. 
\item
Custodial SU(2) violation measured by $T\simeq m_t^2$ agrees 
very well with the \SM value. This strongly disfavours scenarios 
which have sizable extra custodial SU(2) violating effects, 
such as additional fermionic doublets beyond the \SMp 
Unless special care is taken the mechanism responsible for the 
top-bottom mass splitting will lead to huge effects via extra doublets. 
Thus, in order to avoid fine-tunings or special choices, we
avoid extra fermionic doublets beyond the \SMp
\item
The absence of flavour changing neutral currents (FCNC's) beyond the
\SM easily becomes a problem in models where the heavy top mass is 
explained {\em after} the \EW symmetry is broken. 
A well known example is the generation of quark masses in extended 
Technicolor. This might suggest that the heavy top mass 
is intimately related to \EW symmetry breaking, which leads
to the idea of top-condensates. 
\end{itemize}

\noindent
Dynamical symmetry breaking models along these guidelines should be 
phenomenologically much more viable than, e.g., naive Technicolor since 
the model has a \SMLp The above choices are not very artificial and 
the physically most interesting point is probably to understand which 
sort of dynamics produces the scalar spectrum.

\section{The BHL Example}

There is a nice prototype model which implements the \SML in a
minimal way: The so-called BHL model~\cite{BHL} of \EW
symmetry breaking. The idea is here to eliminate the fundamental Higgs
field in the \SM and to introduce instead a new attractive
interaction, which leads to the formation of a
$t\bar{t}$-condensate. In terms of the discussion given above, the
elementary Higgs field of the \SM gets replaced by the bi-fermion
composite operator \beq \hat O \simeq \bar Q_Lt_R~,
\label{OP}
\eeq
with exactly the \SMo-Higgs quantum numbers. We found previously that 
scalars are favoured as unitarity partners of the \GBs due 
the constraints on the S parameter. In addition, in this approach the 
generation of the large top mass occurs within the process of gauge 
symmetry breaking. Therefore $m_t$ is naturally of the order of the 
\EW scale and need not be artificially generated at the cost of possibly 
large FCNC's.

The interaction responsible for triggering the condensation is assumed 
to have its origin in new yet unspecified physics above some high-energy 
scale $\Lambda$. At lower energies this sector of new physics manifests 
itself through non-renormalizable interactions between the usual fermions 
and gauge bosons, where for energies $E \ll \Lambda$ the lowest dimensional 
four-fermion operators are most important. Thus, at the scale $\Lambda$ 
the Lagrangian can be given by the gauge-kinetic terms for the known chiral 
fermions and gauge bosons, plus a gauge invariant four-fermion interaction 
term
\beq
{\cal L} = {\cal L}_{kin}(g,f) + {\cal L}_{4f}^{BHL}~.
\label{L4f1}
\eeq
The gauge invariant four-fermion operator which is 
needed for a condensation of the 
operator $\hat{O}$ [see~(\ref{OP})] in the $t\bar{t}$-channel is 
given by~\cite{N,Mir,BHL,Mar} 
\beq
{\cal L}_{4f}^{BHL}=G(\bar{Q}_{Li}t_R)(\bar{t}_R Q_{Li})~.
\label{4f1}
\eeq
Here $Q_L$ is the left-handed doublet of the third generation quarks, 
$G$ is a dimensionful coupling constant, $G \sim \Lambda^{-2}$, and it 
is implied that the color indices are summed over within each bracket.

The above Lagrangian represents a gauged NJL-model, where condensation 
and \EW symmetry breaking can occur for $G>G_{critical}$. The 
model can be studied analytically in the large $N_c$ (number of colors) 
limit in the so-called NJL or fermion bubble approximation~\footnote{We 
use the well known abbreviation NJL though the paper of Vaks and Larkin 
was received and published first.}$^,$~\cite{VL,NJL}. 
In auxiliary field formalism one can define the local composite operator 
$\phi := -G\bar t_RQ_L$, which allows the Lagrangian~(\ref{L4f1}) to be 
rewritten with the help of the equations of motion into 
\beq
{\cal L} = {\cal L}_{kin}(g,f) 
- \bar Q_L \phi t_R - \bar t_R\phi^+Q_L -G^{-1}\phi^+\phi~.
\label{Laux1}
\eeq
Integrating out the degrees of freedom between $\mu < \Lambda$ and 
$\Lambda$ radiatively generates the additional renormalizable 
terms~\footnote{For $\Lambda \simeq v$ many non-renormalizable 
terms would be generated as well.} for~(\ref{Laux1}). The resulting 
effective Lagrangian at $\mu << \Lambda$ can be written as
\bea
{\cal L}_{\rm eff} & = &
{\cal L}_{kin}(g,f) + \delta{\cal L}_{kin}(g,f) \nonumber \\
& & + Z_\phi |D\phi|^2 
- (1+\delta g_t) \left(\bar Q_L \phi t_R + \bar t_R\phi^+Q_L\right) 
\nonumber \\
& & - (G^{-1} - \delta M^2)\phi^+\phi 
- \frac{\delta\lambda}{2}(\phi^+\phi)^2 ~.
\label{Leff1}
\eea
This is the effective Lagrangian for the composite operator 
$\hat O \equiv \phi$, and the CJT effective potential~\cite{CJT} 
in the fermion bubble approximation can easily be read off:
\beq
V_{\rm eff}(\phi)=(G^{-1} - \delta M^2)\phi^+\phi 
+ \frac{\delta\lambda}{2}(\phi^+\phi)^2 ~.
\label{Veff1}
\eeq
The terms $\delta M^2$ and $\delta \lambda$ follow from
the one loop diagrams with two and four external composite
operators connected via four-fermion (effectively Yukawa)
vertices with weight $G^{-1}$.
The potential (\ref{Veff1}) leads to symmetry breaking for 
small $G^{-1}$, i.e. large enough $G>G_{critical}$. 
Up to the unconventionally normalized kinetic term, 
the $\phi$-sector in the effective Lagrangian~(\ref{Leff1})
looks like the Higgs sector of the \SMp  The difference is 
that here the Higgs is composite, and the scale dependence 
on the infrared cutoff $\mu$ must be such that all the quantum 
effects generated in~(\ref{Leff1}) disappear at $\mu=\Lambda$.
These are the so-called ``compositeness conditions'' 
\beq
Z_\phi 
\stackrel{\mu^2\rightarrow\Lambda^2}{\longrightarrow} 0~;
\quad
\delta M^2
\stackrel{\mu^2\rightarrow\Lambda^2}{\longrightarrow} 0~;
\quad
\delta\lambda
\stackrel{\mu^2\rightarrow\Lambda^2}{\longrightarrow} 0~.
\label{cond}
\eeq
The rescaling $\phi\longrightarrow\phi/\sqrt{Z_\phi}$ 
normalizes the kinetic term for the composite Higgs field 
to unity. Rewriting the compositeness conditions in this 
normalization one finds the boundary conditions which the \SM must 
fulfill if the Higgs particle emerges from top condensation. 

These boundary conditions constitute restrictions on the parameter
space of the low energy effective theory, and therefore lead to
predictions which can be verified against experimental data.  In the
fermion bubble approximation the symmetry breaking and mass generation
is calculable by minimizing the effective potential in
equation~(\ref{Veff1}).  This procedure is equivalent to computing a
self-consistent dynamical top mass by solving the gap equation and
yields a prediction of $m_t$ in terms of the $W$-boson (see section
5), which depends logarithmically on the scale of new physics
$\Lambda$.  In addition one obtains (in bubble approximation) the 
relation $m_H = 2 m_t$ for the Higgs boson mass, which is very suggestive 
for the Higgs being a $\bar{t}t$ bound-state. For 
$\Lambda \approx 10^{15}$~GeV one finds a value of $m_t \approx 165$~GeV, 
while $\Lambda\simeq$ TeV leads to a top mass of a few hundred GeV.  
To obtain a phenomenologically acceptable value for the top quark mass
one has to tune the four-fermion coupling constant $G$ extremely close
to its critical value to allow $m_t/\Lambda$ to be tiny. It has been
shown~\cite{BHL} that this is equivalent to the usual fine-tuning of
the Higgs boson mass in the \SMp Thus, the gauge hierarchy problem is
not solved in the top-condensate approach~\footnote{It has been
claimed in~\cite{Bl} that taking into account the loops with composite
Higgs scalars results in the automatic cancellation of quadratic
divergences. Here we do not discuss this possibility further.}.

However, the bubble approximation employed so far is a rather 
crude approximation of the full dynamics and ignores important 
effects of the full theory, such as QCD corrections and the 
contributions of propagating composite scalars. An elegant way 
to incorporate these effects is to impose the above boundary 
conditions~(\ref{cond}) on the \RG flow (RGE) of the full 
low energy theory, the \SMp  The top mass prediction is then 
governed by the so-called infrared quasi fixed-points~\cite{IRqFP} 
of the top Yukawa coupling. Due to the focusing one obtains 
rather reliable predictions, mostly independent of non-perturbative 
effects close to the cutoff. It turns out that the improved top 
mass predictions are systematically higher than in bubble 
approximation. For a desirable low value for the cutoff in the 
TeV-range the top mass is too large by a factor of two. Even for 
undesirably large scales of $\Lambda > 10^{15}$~GeV one finds 
values around $m_t = 220-240$~GeV~\cite{BHL}, which are about 
30 percent too high. Therefore, the BHL model is phenomenologically 
unacceptable due to an embedding relic: The top mass prediction.

Even though the BHL model is ruled out by the top mass prediction 
it demonstrates how the decoupling of physics beyond the \SM can 
be achieved in \DSB models. 
At scales far below the scale of new physics $\Lambda$, the 
effective theory features a symmetry breaking sector with 
a scalar (composite) boson with exactly the quantum numbers of the 
\SM Higgs boson. From the low energy point of view, the model is
indistinguishable from the \SMk i.e., a \SMLk and the only remnant is
the prediction for the top mass. It is therefore interesting to see 
which kind of theories at higher energies could justify the
effective non-renormalizable four-fermion interactions of the BHL
model with a cutoff. The four-fermion term of the model changes 
under Fierz transformation into the remarkable simple structure:
\beq
G\bar{L}t_R\bar{t_R}L 
\stackrel{Fierz}{\longrightarrow}
-\frac{G}{2} \left(\bar{L}\gamma_\mu L\right)
\left(\bar{t}_R\gamma_\mu t_R\right)~.
\label{fierz}
\eeq 
It is now easy to see that the four-fermion structure of the BHL
model may be related to the exchange of suitable massive, strongly
coupled vector bosons. Thus, such a scenario might be justified
within a renormalizable theory with extended gauge group where the
massive propagator has been integrated out. A number of renormalizable
gauge theory models have been proposed along this line~\cite
{topcolor,topcolor2,U1,SU2V}. Note that the dynamics of such scenarios
deviates clearly from QCD in this picture.  There are even hints for
interesting confinement-Higgs dualities which might play a role in
such models~\cite{topcolor2}.

\section{More Condensates}

Most attempts to modify the BHL model have a common problem: They tend
to produce a top mass which is unacceptably high, even for very high
values for the scale of new physics.  We will see that this is not an
accident, but occurs systematically in scenarios of dynamical \EW 
symmetry breaking which is driven by a top condensate alone. If a top 
condensate breaks the \EW symmetry, then for an asymptotically free 
theory the dynamically generated top propagator can be written as
\beq S_t =
\frac{i}{\slash{p}-\Sigma_t(p^2)} ~,
\label{St}
\eeq
where 
\beq
\Sigma(p^2)\stackrel{p^2\rightarrow\infty}{\longrightarrow}0~.
\label{ass}
\eeq
The dynamically generated self energy $\Sigma_t$ can be related to the
\GB decay constant $F_\pm$, which is induced by the condensate, 
by virtue of the so-called Pagels--Stokar relations~\cite{PS}: 
\beq F_\pm^2 = \frac{N_c}{32\pi^2} \int dk^2
\frac{\Sigma^2_t(k^2)}{(k^2-\Sigma^2_t(k^2))}~.
\label{PS}
\eeq
These relations are derived primarily on the basis of Ward identities and 
are therefore valid beyond the approximations used to derive $\Sigma_t(p)$. 
Note that the integral in~(\ref{PS}) is formally log
divergent, but finite if the asymptotic behaviour of equation~(\ref{ass}) is
taken into account. Equation~(\ref{PS}) is a very powerful relation between
the dynamically generated top mass and the \GB decay constant, and
also the W mass, since after the charged \GB is absorbed by $W$ one
obtains $m_W^2=g_2^2 F_\pm^2$. It is instructive to observe
that the integral on the right hand side of the Pagels--Stokar relation,
equation~(\ref{PS}), feels the structure of $\Sigma_t$ only on a
logarithmic scale. Thus, for large values of $\Lambda$, the integral is 
dominated by contributions coming from scales both far away from
$\Lambda$ and the \EW scale, i.e.  from regions where
$\Sigma$ should (up to logarithmic RGE running) essentially be
flat. In this case it is a good approximation to express $\Sigma$
by its `mean' height and extension: 
\beq \Sigma_t(p^2) = m_t
\Theta(\Lambda^2-p^2)~.
\label{ansatz}
\eeq 
To be more precise, this should be a valid approximation {\em except} 
for small values of $\Lambda$, but those values anyway lead to values 
of $m_t$ that are by far too large.  Inserting equation~(\ref{ansatz}) 
into equation~(\ref{PS}) and solving for the top mass, one finds

\beq m_t^2 = \frac{32\pi^2
m_W^2}{N_c~g_2^2~\ln(\Lambda^2/m_t^2)}~,
\label{PSmt}
\eeq
which is exactly the relation obtained in the BHL model in bubble 
approximation. This makes sense, since the structure of the 
ansatz~(\ref{ansatz}) corresponds to a \NJL gap equation. Corrections 
to this relation come, like in the BHL model, from other, weak gauge 
contributions and are expected to be moderate and model dependent. 
Thus, the Pagels--Stokar relation explains why many variants of the 
BHL model, like two Higgs doublets or the supersymmetric version, 
produce a similar top mass, a value too high to fit the data.
On the other hand, inserting $m_t=175$~GeV in equation~(\ref{PSmt}) 
one finds a $W$ mass which is too small for any $\Lambda$. Therefore, 
in order to obtain a viable relation between $\Lambda$ and the top mass, 
one is led to consider more complex symmetry breaking scenarios, 
with more condensates and/or more Yukawa couplings.

Given the success of renormalizable gauge theories, it appears 
attractive to relate extra condensates to more complex symmetry 
breaking patterns of extended gauge sectors. An appealing class 
of models where such ideas can be exemplified are dynamically broken 
\LRS theories. In such models, usually a condensate in the 
leptonic sector breaks \LR symmetry as a first step, and a 
second (or more) condensate(s) cause \EW symmetry breaking. 
We will show in the next section how \DSB can 
be implemented in \LRS theories, along the guidelines 
presented in section 3, to give a model with a \SMLp In particular, 
we will find a modified relation between $m_t$ and $\Lambda$ which is 
phenomenologically acceptable - even for low values of $\Lambda$. 

\section{A Left-Right-Symmetric Model with Decoupling Limit}

Left-right-symmetric models based on the gauge group 
$\mbox{SU(2)}_L\times \mbox{SU(2)}_R\times \mbox{U(1)}_{B-L}\times 
\mbox{SU(3)}_c$ 
have many attractive features~\cite{LR1,LR3}. The U(1)--group 
corresponds to baryon minus lepton number and thus has a somewhat more 
intuitive interpretation than the weak hypercharge of the \SMp 
Furthermore, parity violation is explained as yet another step of 
spontaneous gauge symmetry breaking. All known quarks and leptons fit 
nicely in the fundamental representation of the gauge group, and the 
well known see-saw mechanism can be employed to obtain very light masses 
for the left-handed neutrinos. 

The symmetry breaking sequence required by phenomenology proceeds 
in two steps, first the breaking of parity down to the \SM gauge group at 
an energy scale $\mu_R$, and second the usual \EW symmetry breaking:
\beq
\begin{array}{c}
  \mbox{SU(2)}_L \otimes \mbox{SU(2)}_R \otimes \mbox{U(1)}_{B-L} \\
  \downarrow{\scriptstyle \mu_R\hspace{0.27em}}   \\
  \mbox{SU(2)}_L \otimes \mbox{U(1)}_{Y}~ \\
  \downarrow{\scriptstyle \mu_{ew}}\\
  \mbox{U(1)}_{em} 
\end{array} \label{LRewsymm}
\eeq
In conventional \LRS models this breaking sequence is realized by 
introducing an elementary scalar Higgs sector. Naturally in this case 
the scalar sector is much larger than in the \SMk with of the order 
of twenty parameters to be adjusted to obtain the desired symmetry 
breaking pattern. In the light of the above considerations,
it will be interesting to investigate whether the above symmetry 
breaking scenario can be realized in a dynamical model with composite 
Higgs bosons.  

In the following we discuss composite model building in the framework 
of \LR symmetric theories, in analogy to the BHL model. We will show 
that the compositeness of scalars poses non-trivial constraints on 
the models that can be realized, and finally present a phenomenological 
viable \LRS model, where the correct vacuum structure and symmetry 
breaking sequence emerges dynamically~\cite{ALSV1,ALSV2}. Essentially 
the symmetry breaking in this composite model involves at least two 
condensates, one large hybrid condensate in the neutrino sector to 
break parity, and, in the simplest case, a $t\bar{t}$-condensate to 
break the \EW symmetry. As it turns out, the general case gives 
three condensates, two of which will be in the quark sector
and appear as the two vacuum expectation values 
of the bi-doublet composite scalar $\phi$. 
These two condensates break the \EW gauge symmetry and, as 
we pointed out in the last section,  provide the necessary degree of 
freedom to obtain viable top and bottom masses. 

\subsection{A First Attempt}

As in the BHL approach, we will only consider the usual fermions and 
gauge bosons of the model as elementary particles, with no fundamental 
Higgs scalars being present, and in addition introduce a set of relevant 
\LRS four-fermion interactions, representing yet unspecified new physics 
at a high energy scale $\Lambda$. Thus the Lagrangian will be of the form
\beq
{\cal L} = {\cal L}_{kinetic}^{LR} + {\cal L}_{4f}~.
\label{L}
\eeq
A useful framework to study models with composite Higgs bosons is the
auxiliary field technique. The four-fermion 
interactions are rewritten in terms 
of Yukawa couplings of the fermions with newly introduced static auxiliary 
fields with appropriate quantum numbers. The non-propagating scalars come 
with a heavy mass term of the order of $G_{4F}^{-1}\approx \Lambda^2$, but 
no kinetic term and no quartic interactions on tree level. Since the 
modified Lagrangian of the system is quadratic in these auxiliary fields 
they can always be integrated out in the functional integral~\cite{Auxform}. 
Equivalently, one can use the equations of motion for these fields to express 
them in terms of the fermionic degrees of freedom. After substituting the 
resulting expressions into the auxiliary Lagrangian one reproduces the 
initial four-fermion structures.
The static auxiliary fields can acquire gauge-invariant kinetic terms and
quartic self-interactions through radiative corrections and become
physical propagating scalar fields at low energies provided that the
corresponding gap equations are satisfied~\cite{BHL}. The kinetic terms and
mass corrections can be derived from the 2-point Green function, whereas
the quartic couplings are given by the 4-point functions. Given the Yukawa
couplings of the scalar fields one can readily calculate these functions in
the fermion bubble approximation, in which they are given by the
corresponding 1-fermion-loop diagrams.

Before presenting the final model, we will further consider the type 
of composite scalars we wish to obtain, and the constraints imposed 
by a correct symmetry breaking sequence. 
In the most popular \LRS model the Higgs sector consists of a
bi-doublet $\phi \sim (2,2,0)$ and two triplets, $\Delta_L \sim (3,1,2)$
and $\Delta_R \sim (1,3,2)$, representations in parentheses corresponding to 
$\mbox{SU(2)}_L \otimes \mbox{SU(2)}_R \otimes \mbox{U(1)}_{B-L}$.
Assuming that these scalars are bound-states of the usual fermions, the 
following fermionic content reproduces the correct quantum numbers:
$$\phi_{ij} \sim \alpha (\bar{Q}_{Rj}Q_{Li}) + \beta (\tau_2 \bar{Q}_L Q_R
\tau_2)_{ij}+\mbox{leptonic terms}~, $$
\beq
\vec{\Delta}_L \sim (\Psi_L^T C \tau_2 \vec{\tau}\Psi_L),\;\;
\vec{\Delta}_R\sim (\Psi_R^T C  \tau_2 \vec{\tau}\Psi_R)~.
\label{phiij}
\eeq
Here $Q_L,\Psi_L$ ($Q_R,\Psi_R$) are left-handed (right-handed) doublets
of quarks and leptons, respectively; $i$ and $j$ are isospin indices.
In models with composite Higgs bosons generated by four-fermion operators 
the scalars are, roughly speaking, ``square roots'' of these four-fermion 
operators due to the equations of motion. A good starting point to find a 
set of four-fermion-operators leading to the desired scalars~(\ref{phiij}) 
is to square these expressions. 

However, the model with composite triplet scalars is not going to be 
viable: The dynamically generated effective potential is constrained and 
does not allow the phenomenologically required spontaneous parity breaking. 
The following little exercise will help us to subsequently construct a viable 
model. Consider spontaneous parity breakdown in \LR models with composite Higgs
bosons. It is usually assumed that, in addition to the gauge symmetry, the
Lagrangian of the \LR model possesses a discrete parity symmetry under which
\beq
Q_L \leftrightarrow Q_R,\;\;\; \Psi_L \leftrightarrow
\Psi_R,\;\;\; \phi \leftrightarrow \phi^\dagger,\;\;\;
\Delta_L \leftrightarrow \Delta_R,\;\;\;W_L\leftrightarrow W_R~.
\label{discrete}
\eeq
Then the $\Delta$-sector of the (radiatively
induced) scalar potential contains the terms 
\bea
V(\Delta_L,\Delta_R) & = &
     - m^2 (\Delta_L^\dagger\Delta_L+\Delta_R^\dagger \Delta_R ) \\
&& +\lambda_1 [(\Delta_L^\dagger \Delta_L)^2+(\Delta_R^\dagger \Delta_R)^2]\\
&& +2\lambda_2 (\Delta_L^\dagger \Delta_L)(\Delta_R^\dagger \Delta_R)
+ \ldots~,
\label{Vtrip}
\eea
Despite the discrete parity symmetry, parity can be spontaneously broken 
if $\langle \Delta_R \rangle >  \langle \Delta_L \rangle$~\cite{LR2}.
From~(\ref{Vtrip}) this can only happen for $\lambda_2 >\lambda_1$, and 
in the conventional approach $\lambda_1$ and $\lambda_2$ are chosen 
appropriately as free parameters of the model. However, the scalar mass 
terms and couplings in the composite Higgs approach are not arbitrary; 
they are all calculable in terms of the four-fermion couplings $G_a$ and
the scale of new physics $\Lambda$~\cite{BHL}. In particular, in the fermion
bubble approximation at one loop level the quartic couplings $\lambda_1$ and
$\lambda_2$ are induced through the Majorana-like Yukawa couplings
$f(\Psi_L^T C \tau_2 \vec{\tau}\vec{\Delta}_L \Psi_L+\Psi_R^T C\tau_2
\vec{\tau}\vec{\Delta}_R \Psi_R)+h.c.$, and are given by the diagrams of
figure~1.
\begin{figure}[htb]
\centerline{
\rotate[r]{\epsfxsize=27ex \epsffile{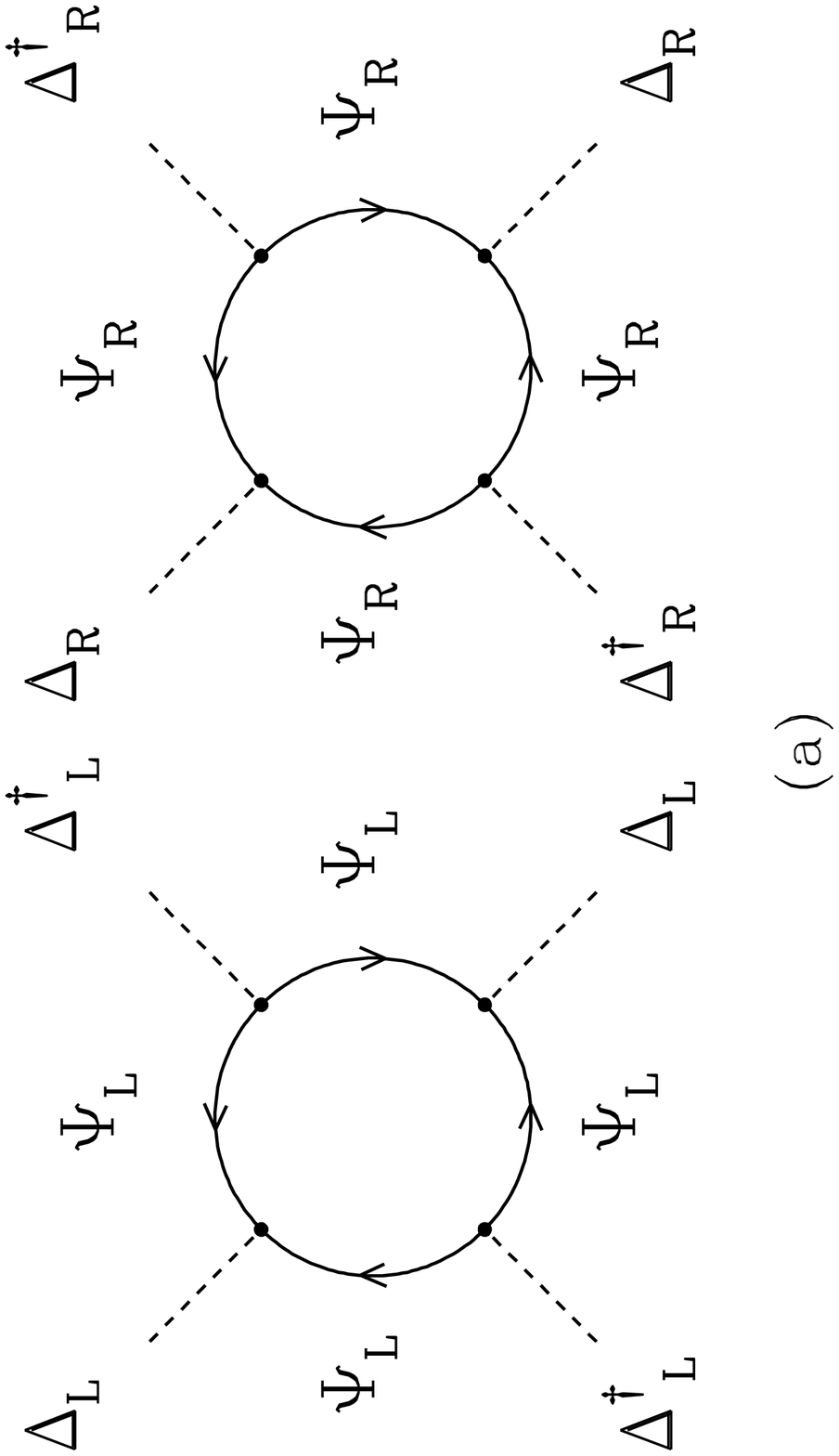} }
\hspace{2em}
\rotate[r]{\epsfxsize=27ex \epsffile{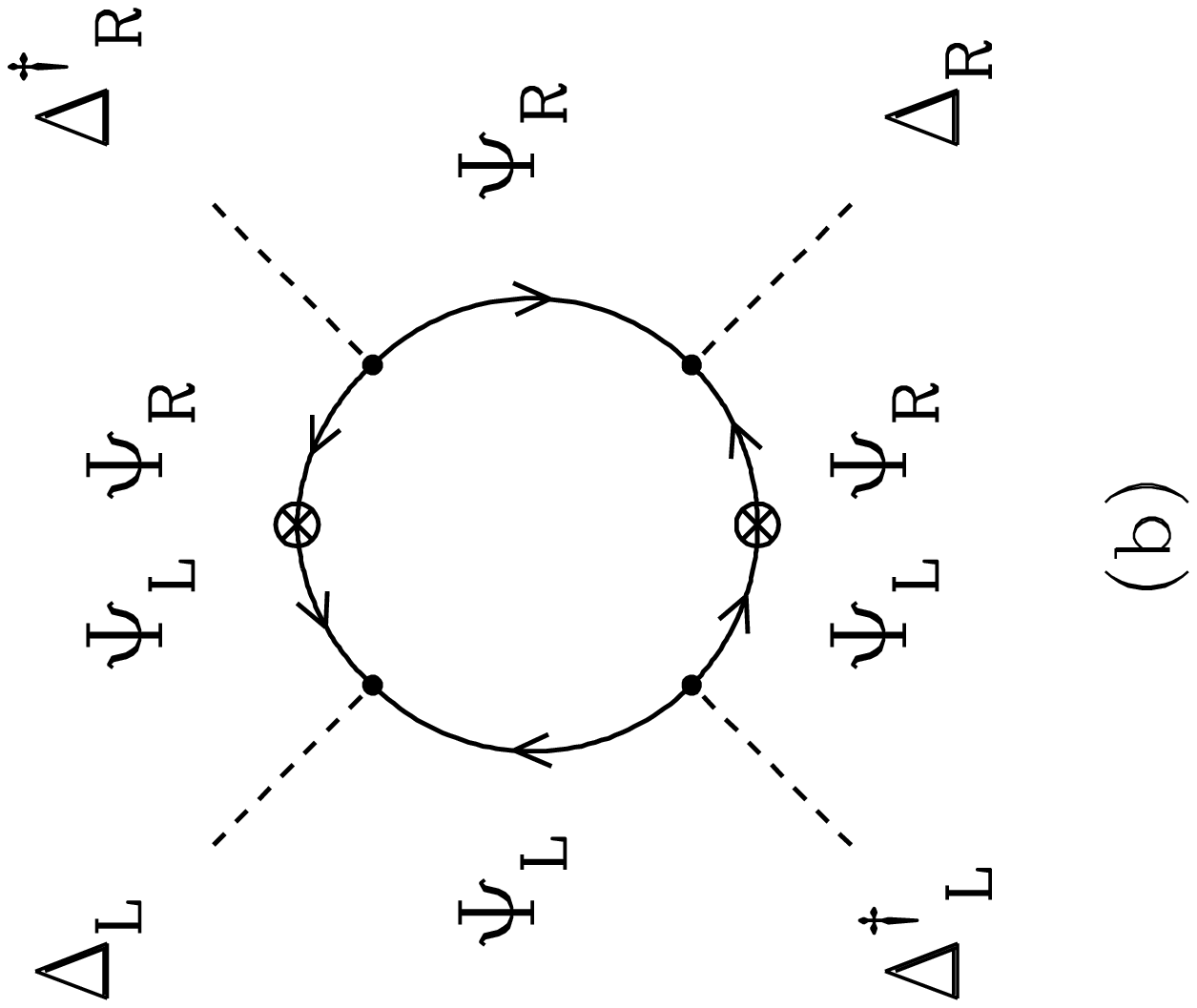} }
}
\caption[]{\small\sl
Fermion loop diagrams contributing to the quartic couplings
$\lambda_1$~(figure~\ref{fig:delta}a) and 
$\lambda_2$~(figure~\ref{fig:delta}b)
for Higgs triplets.
\label{fig:delta}
}
\end{figure}
It can be seen from figure~1b that to induce the $\lambda_2$ term one 
needs the $\Psi_L$--$\Psi_R$ mixing in the fermion line in the loop, 
i.e. the lepton Dirac mass term insertions. However, the Dirac mass terms 
are generated, e.g., by the vacuum expectation values 
of the bi-doublet $\phi$; they are absent 
at the parity breaking scale which is supposed to be higher than the \EW 
scale. Even if parity and the \EW symmetry are broken simultaneously 
(which is hardly a phenomenologically
viable scenario), this would not save the situation since the diagram of
figure~1b is finite in the limit $\Lambda\to\infty$ whereas the one of 
figure~1a is logarithmically divergent. Consequently, the inequality 
$\lambda_2>\lambda_1$ cannot be satisfied.

\subsection{A Viable Model}

\noindent From the above considerations follows that we have to look for
a modified low energy scalar sector.
Originally \LRS models were constructed including two 
doublets~\cite{LR1}, $\chi_L \sim (2,1,-1)$ and $\chi_R \sim (1,2,-1)$.
However, these cannot have any gauge invariant interactions with 
the known quarks and leptons, while the subsequent models with triplets
$\Delta_L$ and $\Delta_R$ are attractive due to Majorana like couplings
to leptons and the resulting see-saw mechanism. 

In the composite approach the disadvantage of the doublet-model can be 
turned into a virtue, the only ``price'' being the introduction of a new 
gauge singlet fermion. A gauge singlet fermion is required to
construct a composite scalar doublet out of fermion bi-linears, since all 
the known fermions already come in doublets. However, this singlet is 
quite welcome, since its Majorana like coupling leads to a modified see-saw
mechanism~\cite{WW}, which can naturally explain the smallness of 
neutrino masses. Second, we will see that the chiral singlet fermion
in the loops is essential for the correct \LR symmetry breaking pattern to 
emerge in the composite doublet model.

We therefore assume that in addition to the usual quark and lepton doublets 
there is a gauge-singlet fermion
\beq
S_L \sim (1,\;1,\;0)~.
\label{SL}
\eeq
To maintain the discrete parity symmetry one needs a right-handed 
counterpart of $S_L$. This can be either another particle, $S_R$, or the 
right-handed antiparticle of $S_L$, $(S_L)^c\equiv C\bar{S}_L^T = S^c_R$. 
The latter choice is more economical and at the same time crucial for the 
$\lambda_2$ contribution of the effective potential and the desired 
symmetry breaking pattern. We therefore assume that under parity operation
\beq
S_L \leftrightarrow S^c_R~.
\label{SLC}
\eeq
With this new singlet and the usual quark and lepton doublets we introduce
the following set of gauge-invariant four-fermion interactions:
\bea
{\cal L}_{4f}=G_1(\bar{Q}_{Li}Q_{Rj})(\bar{Q}_{Rj}Q_{Li})+
[G_2(\bar{Q}_{Li}Q_{Rj})(\tau_2\bar{Q}_{L}Q_{R}\tau_2)_{ij}
+h.c.]\nonumber \\
+G_3(\bar{\Psi}_{Li}\Psi_{Rj})(\bar{\Psi}_{Rj}\Psi_{Li})+
[G_4(\bar{\Psi}_{Li}\Psi_{Rj})(\tau_2\bar{\Psi}_{L}\Psi_{R}\tau_2)_{ij}
+h.c.]\nonumber \\
+[G_5(\bar{Q}_{Li}Q_{Rj})(\bar{\Psi}_{Rj}\Psi_{Li})+h.c.]+
[G_6(\bar{Q}_{Li}Q_{Rj})(\tau_2\bar{\Psi}_{L}\Psi_{R}\tau_2)_{ij}
+h.c.] \nonumber \\
+G_7[(S_L^T C \Psi_L)(\bar{\Psi}_L C \bar{S}_L^T)+
(\bar{S}_L\Psi_R)(\bar{\Psi}_R S_L)]+G_8 (S_L^T C S_L)(\bar{S}_L C
\bar{S}_L^T)~.
\label{L4f}
\eea
In analogy to the BHL model the $G_a$ are dimensionful four-fermion couplings
of the order of $\Lambda^{-2}$ motivated by some new physics at $\Lambda$.
Notice that the above interactions are not only gauge-invariant, but also
(for hermitian $G_2$, $G_4$, $G_5$ and $G_6$) symmetric with respect to the
discrete parity operation (\ref{discrete}), (\ref{SLC}). Note that 
equation~(\ref{L4f}) is a rather general ansatz. It will turn out that 
we need only a subset of the above terms in order to break the 
symmetries correctly.

We assume that only the third generation of fermions contribute to
${\cal L}_{4f}$, i.e., deal with a limit where only the heaviest fermions
are massive, while all the light fermions are considered to be massless.
This appears to be a good starting point from where light fermion masses 
could, e.g.,  be generated radiatively. In addition to the bi-doublet 
$\phi$ of the structure given in equation~(\ref{phiij}), the above 
four-fermion couplings, if critical, can give rise to a pair of composite 
doublets $\chi_L$ and $\chi_R$, and also to a singlet scalar $\sigma$:
\beq
\chi_L \sim S_L^T C \Psi_L, \;\;\;\;  \chi_R \sim \bar{S}_L\Psi_R =
(S^c_R)^T C \Psi_R, \;\;\;\; \sigma \sim \bar{S}_L C \bar{S}_L^T~.
\label{composite}
\eeq
From equations~(\ref{discrete}) and (\ref{SLC}) it follows that under parity
we have $\chi_L \leftrightarrow \chi_R$ and $\sigma \leftrightarrow
\sigma^\dagger$. Switching to the auxiliary field formalism, the scalars
$\chi_L$, $\chi_R$, $\phi$ and $\sigma$ have the following bare mass terms
and Yukawa couplings:
\bea
L_{aux}&=&-M_0^2(\chi_L^\dagger \chi_L+\chi_R^\dagger \chi_R)-M_1^2
\tr{(\phi^\dagger \phi)} \nonumber \\
& & -\frac{M_2^2}{2}\tr{(\phi^\dagger\tilde{\phi}+h.c.)}
    -M_3^2 \sigma^\dagger\sigma \nonumber \\
& & -\left[\bar{Q}_L(Y_1\phi+Y_2\tilde{\phi})Q_R +
\bar{\Psi}_L(Y_3\phi+Y_4\tilde{\phi})\Psi_R + h.c.\right] \nonumber \\
& &-\left[Y_5(\bar{\Psi}_L \chi_L S^c_R+\bar{\Psi}_R \chi_R S_L)
+Y_6 (S_L^T C S_L)\sigma + h.c.\right]~,
\label{Laux}
\eea
where the field $\tilde{\phi}\equiv \tau_2\phi^*\tau_2$ has the same quantum
numbers as $\phi$: $\tilde{\phi}\sim (2,\;2,\;0)$. By integrating out the
auxiliary scalar fields one can reproduce the four-fermion structures of
equations~(\ref{L4f}) and express the four-fermion 
couplings $G_1,...,G_8$ in terms of the
Yukawa couplings $Y_1,...,Y_6$ and the mass parameters $M_0^2$, $M_1^2$,
$M_2^2$ and $M_3^2$ (explicit formulas can be found in~\cite{ALSV2}).
In components, the scalar multiplets of the model are
\bea
\phi =  \left( \begin{array}{cc}
\phi_1^0 & \phi_2^+ \\
\phi_1^-& \phi_2^0 \end{array} \right)\,,\;\;\;
 &&
\langle \phi \rangle = \left( \begin{array}{cc}
\kappa & 0 \\  0 & \kappa' \end{array} \right)\,,\;\;\;     \nonumber \\
\chi_{L}=\left( \begin{array}{c}
\chi_L^0 \\ \chi_L^-\end{array} \right) \hspace{2.5ex} \,,\;\;\;
 &&
\chi_{R}=\left( \begin{array}{c}
\chi_R^0 \\ \chi_R^-\end{array} \right) ~. 
\label{bidef}
\eea
Let us now consider parity breaking in the present \LR model. In a viable
scenario the $\mbox{SU(2)}_R$ symmetry should be broken at the right-handed 
scale
$\mu_R$ by $\langle \chi_R^0 \rangle = v_R$, and the \EW symmetry has to be
broken at $\mu_{EW}$ by the vacuum expectation values 
of $\phi$ and possibly of $\chi_L^0$
($\equiv v_L$). Using the Yukawa couplings of the doublets $\chi_L$ and
$\chi_R$ [see equation~(\ref{Laux})], one can calculate the fermion-loop
contributions to the  quartic couplings $\lambda_1 [(\chi_L^\dagger\chi_L)^2
+(\chi_R^\dagger \chi_R)^2]$ and $2\lambda_2 (\chi_L^\dagger\chi_L)
(\chi_R^\dagger \chi_R)$ in the effective Higgs potential (figures~2a and 2b).
\begin{figure}[htb]
\centerline{
\rotate[r]{\epsfxsize=27ex \epsffile{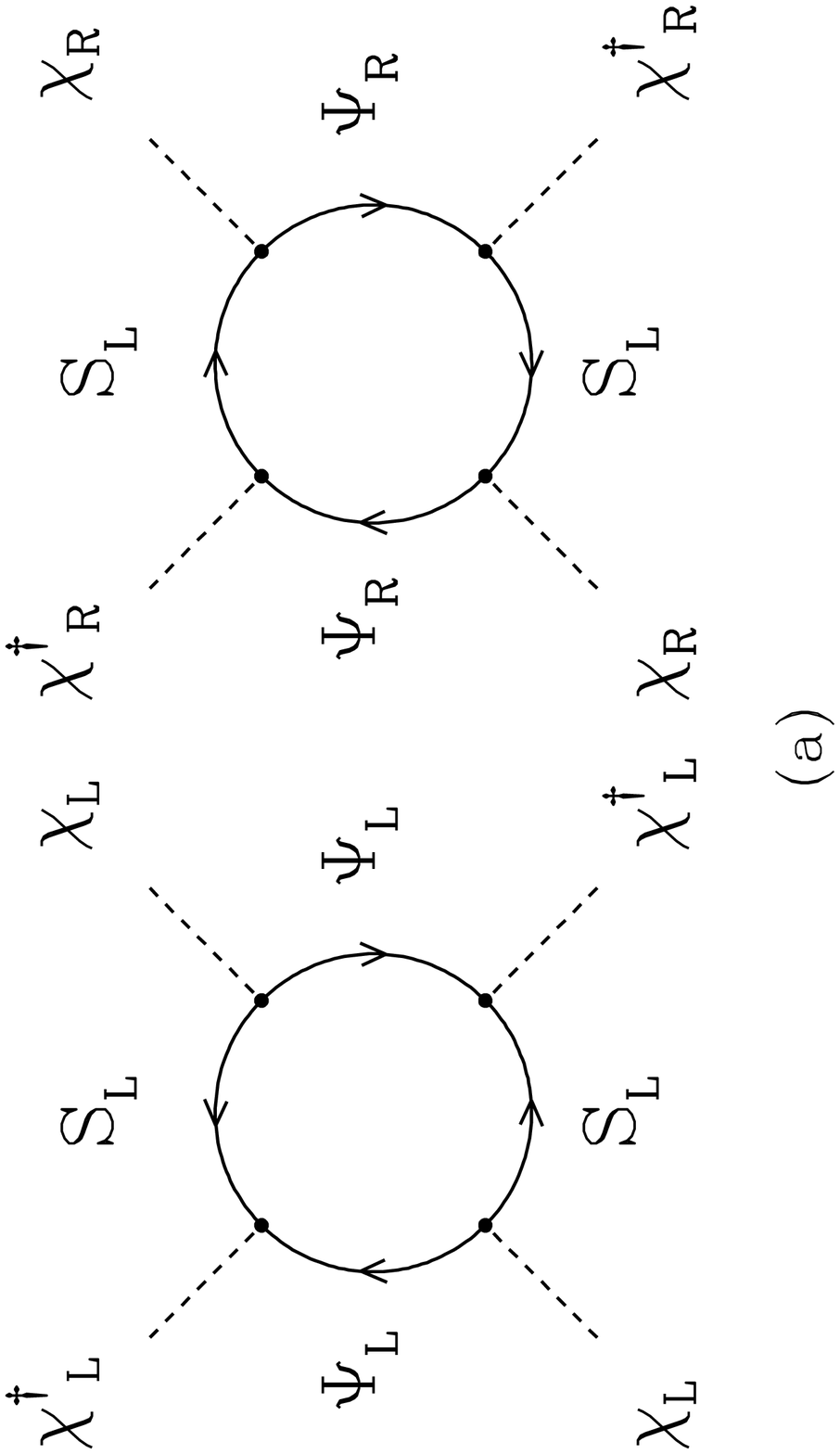} }
\hspace{2em}
\rotate[r]{\epsfxsize=27ex \epsffile{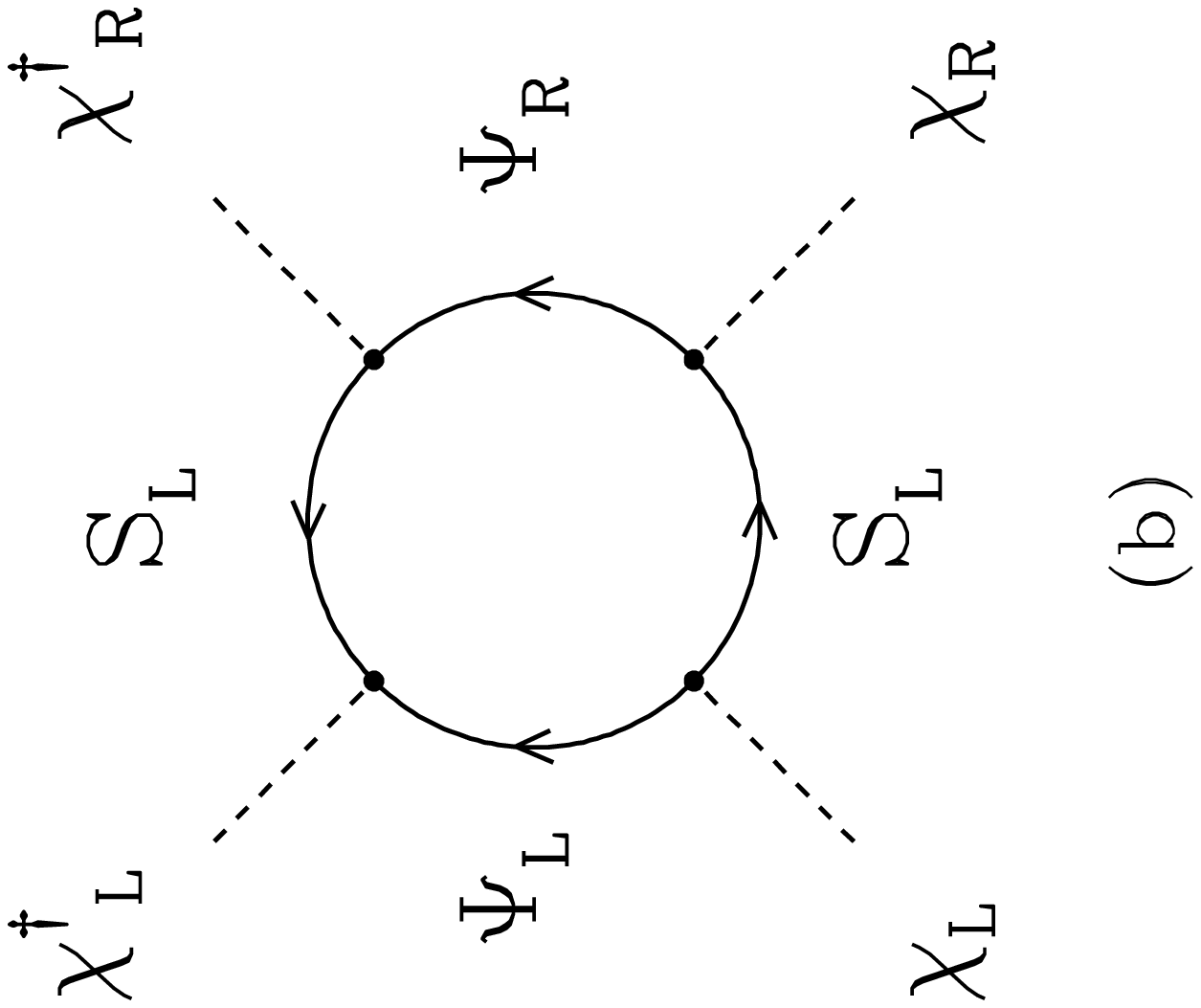} }
}
\caption[]{\small\sl
Fermion loop diagrams contributing to the quartic couplings
$\lambda_1$~(figure~\ref{fig:doub}a) and
$\lambda_2$~(figure~\ref{fig:doub}b) for the Higgs doublets $\chi_{L/R}$.
\label{fig:doub}
}
\end{figure}
Unlike in the triplet case, the $\lambda_1$ and $\lambda_2$ terms are now 
given by similar diagrams, because the gauge singlet couples both to $\chi_L$ 
and $\chi_R$. Since the Yukawa couplings of $\chi_L$ and $\chi_R$ coincide 
(which is just the consequence of the discrete parity symmetry), figures~2a 
and~2b yield $\lambda_1=\lambda_2$. Recall that one needs $\lambda_2> \lambda_1$ 
to have spontaneous parity breakdown, and thus the situation here has 
already improved compared to the triplet scenario. As we shall see,
taking into account the gauge boson loop contributions to $\lambda_1$ and
$\lambda_2$ will automatically secure this relation: Both $\lambda_1$ and 
$\lambda_2$ obtain corrections from $\mbox{U(1)}_{B-L}$ gauge boson loops, 
whereas only $\lambda_1$ is corrected by diagrams with $W^i_L$ or $W^i_R$ 
loops (see figure~3).
\begin{figure}[htb]
\centerline{
\rotate[r]{\epsfxsize=27ex \epsffile{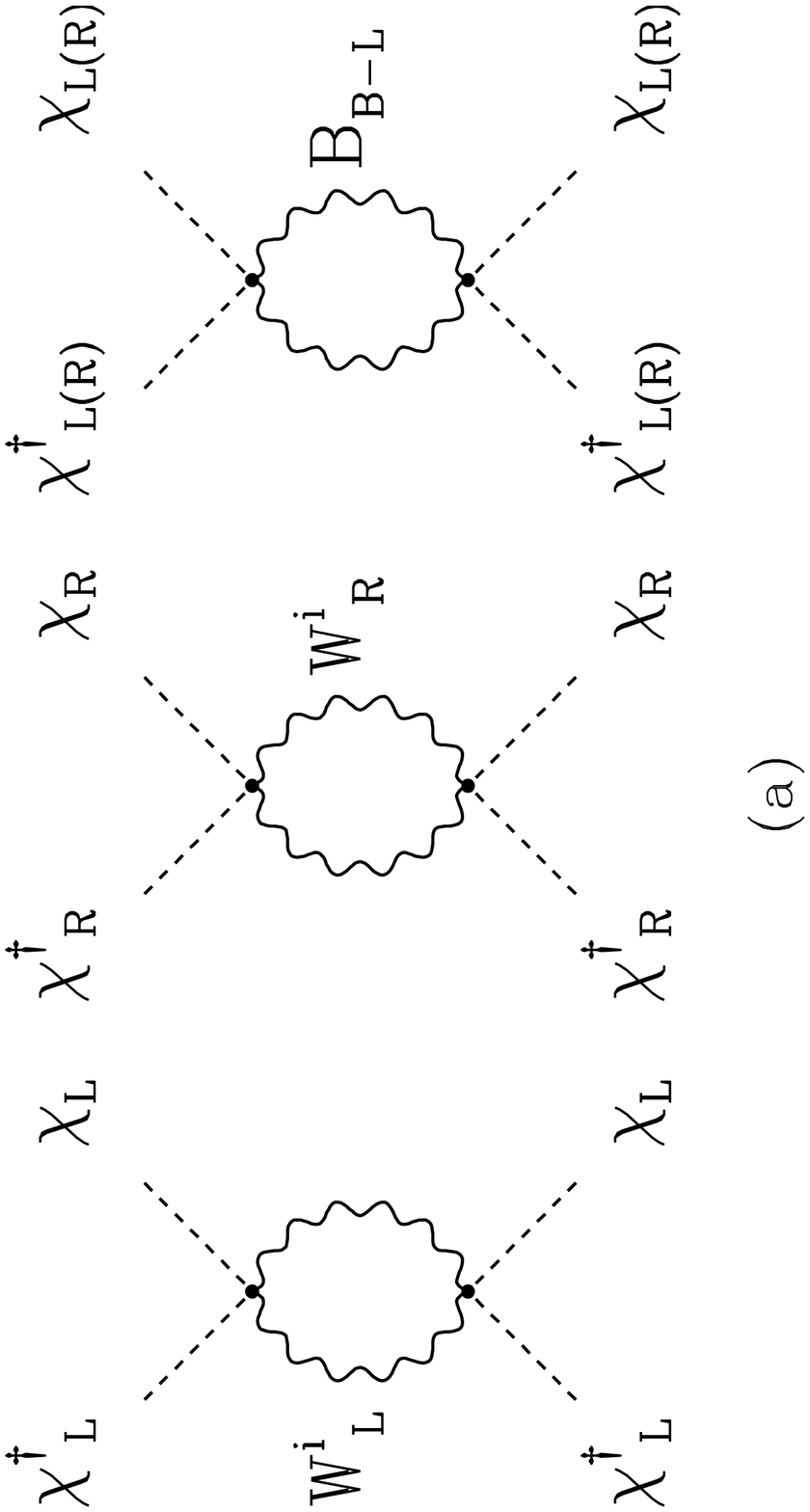} }
\hspace{2em}
\rotate[r]{\epsfxsize=27ex \epsffile{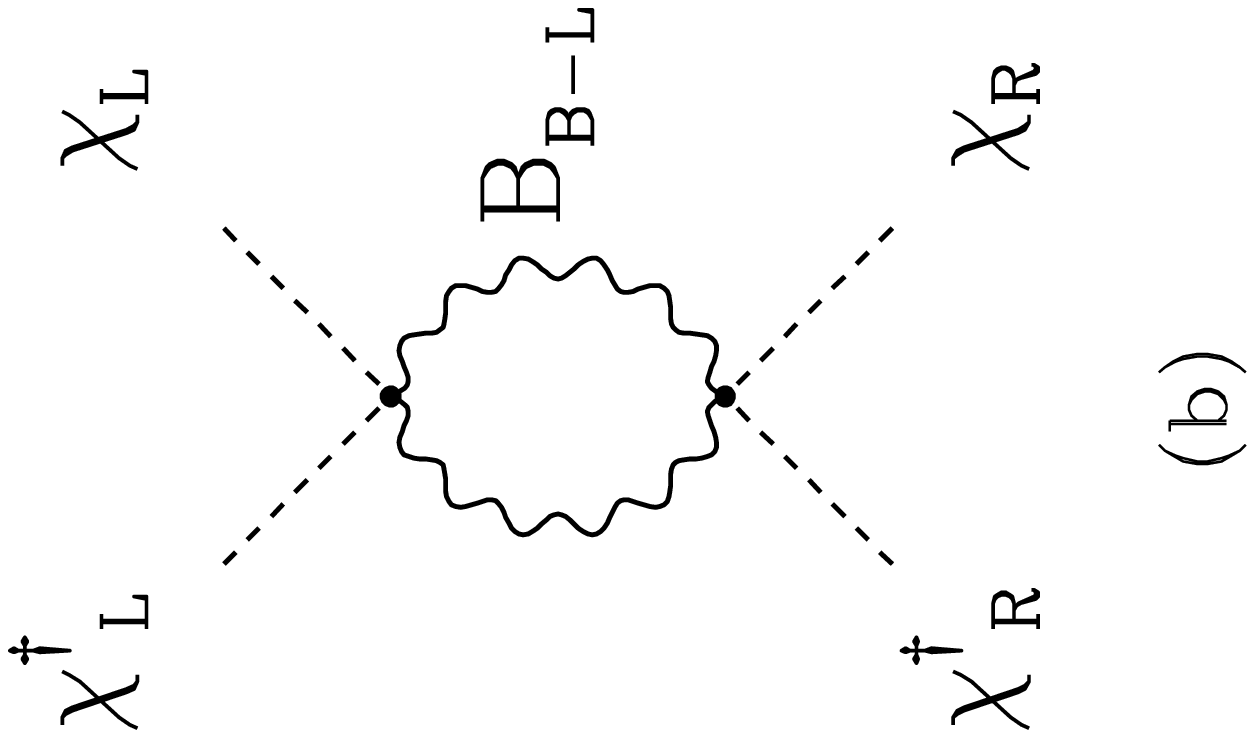} }
}
\caption[]{\small\sl
Gauge boson loop diagrams contributing to the quartic couplings
$\lambda_1$~(figure~\ref{fig:gauge}a) and
$\lambda_2$~(figure~\ref{fig:gauge}b) for the Higgs doublets $\chi_{L/R}$
in Landau gauge.
\label{fig:gauge}
}
\end{figure}
Since all these contributions have a relative minus sign compared to the
fermion loop ones, one finds $\lambda_2>\lambda_1$ irrespective of the
values of the Yukawa or gauge couplings or any other parameter of the model,
provided that the SU(2) gauge coupling $g_2\neq 0$ [compare the expressions
for $\lambda_1$ and $\lambda_2$ in~(\ref{Lambda}) below]. Thus the condition
for spontaneous parity breakdown is automatically satisfied in the model.

We have a very interesting situation here. In a model with composite triplets
$\Delta_L$ and $\Delta_R$ parity is never broken, i.e. the model is not
phenomenologically viable. At the same time, in the model with two composite
doublets $\chi_L$ and $\chi_R$ instead of two triplets (which requires
introduction of an additional singlet fermion $S_L$) parity is broken
automatically. This means that, unlike in conventional \LR models, in the
composite Higgs approach {\em whether or not parity can be spontaneously
broken depends on the particle content of the model rather than on the
choice of the parameters of the Higgs potential}.

As soon as we know the vacuum structure of the model (see below), i.e. 
the different vacuum expectation values for composite scalar fields, 
equation~(\ref{Laux}) yields the expressions for the resulting fermion 
masses. The masses of the quarks and charged leptons and the Dirac 
neutrino mass $m_D$ are given by the vacuum expectation values of the 
bi-doublet~\footnote{We assume all the vacuum expectation values to be real.}:
\bea
m_t=Y_1\kappa+Y_2\kappa',\;\;\;\;\;m_D=Y_3\kappa+Y_4\kappa',\nonumber \\
m_b=Y_1\kappa'+Y_2\kappa,\;\;\;\;\; m_{\tau}=Y_3\kappa'+Y_4\kappa.
\label{fmass}
\eea
It is well known that \LRS models with only doublet Higgs scalars can 
usually not explain small neutrino masses in a natural way. As pointed 
out earlier, introducing the singlet fermion $S_L$ not only provides 
the spontaneous parity breaking in the composite model, but also cures 
the neutrino mass problem.  It was first noticed by Wolfenstein and 
Wyler~\cite{WW} that with an additional singlet neutral fermion $S_L$ the 
neutrino mass matrix takes the form (in the basis $(\nu_L, \nu^c_L, S_L)$)
\beq
M_{\nu} = \left(
\begin{array}{c}
\end{array}
\begin{array}{ccc}
0 & m_D & \beta\\
m_D & 0 & M_{\nu_R}\\
\beta & M_{\nu_R} & \tilde{\mu}
\end{array} \right)\;,
\label{numass1}
\eeq
where the entries $\beta$, $M_{\nu_R}$ and $\tilde{\mu}$ can be read off from
equation~(\ref{Laux}),
\beq
\beta = Y_5 v_L \, , \quad M_{\nu_R} = Y_5 v_R \, 
\quad  \tilde{\mu} = 2Y_6\sigma_0~,
\label{yukawas}
\eeq
with $\sigma_0\equiv \langle\sigma\rangle$. For $v_R \gg \kappa, \kappa',
v_L$ and $v_R \gta \sigma_0$ one obtains two heavy Majorana neutrino 
mass eigenstates  with the masses $\sim M_{\nu_R}$ and a light Majorana 
neutrino with the mass 
$m_\nu \simeq \tilde{\mu}(m_D^2/M_{\nu_R}^2)-2\beta m_D/M_{\nu_R}$
which vanishes in the limit $M_{\nu_R} \to \infty$. This is the modified 
seesaw mechanism which provides the smallness of neutrino mass.

\subsection{Effective Potential and Vacuum Structure}

The rough analysis of parity breakdown given above demonstrated that 
this model can in principle work. Now the next step is to calculate the 
full low energy effective potential of the theory, analyse the 
vacuum structure and the resulting patterns of spontaneous symmetry
breaking. As we have already seen, this is a nontrivial issue since 
the potential parameters are not free but constrained and calculable 
in terms of the initial 6 Yukawa couplings, or four-fermion-terms, 
respectively. 

The effective potential arises as part of a complete renormalizable 
low energy effective Lagrangian which emerges after integrating out
the high energy degrees of freedom. Explicitly, one analyses the 
scalar two- and four-point functions and finds the effective Lagrangian
\bea
{\cal L}_{\rm eff} &=& {\cal L}_0+{\cal L}_{\rm Yuk}+Z_\phi \tr \[(D^\mu 
\phi)^\dagger
(D_\mu \phi)\] +Z_\sigma (\partial^\mu \sigma)^\dagger (\partial_\mu \sigma) 
\nonumber \\
 & & +Z_\chi\[(D^\mu \chi_L)^\dagger (D_\mu \chi_L)+
(D^\mu \chi_R)^\dagger (D_\mu \chi_R)\]   \\
 & & + V_{\rm eff}~. \nonumber
\label{Leff}
\eea
${\cal L}_0$ contains the gauge-invariant kinetic terms of fermions
and gauge bosons, while ${\cal L}_{\rm Yuk}$ represents the Yukawa-coupling
terms in equation~(\ref{Laux}). The additional terms consist of 
mass corrections, gauge invariant kinetic terms for the scalars and
the full effective scalar potential. 
The effective Higgs potential $V_{\rm eff}$ in equation~(\ref{Leff}) reads
\bea
V_{\rm eff}&= &\tilde{M}_0^2(\chi_L^\dagger\chi_L+\chi_R^\dagger\chi_R)
          +\tilde{M}_1^2 \tr{(\phi^\dagger \phi)}+\frac{\tilde{M}_2^2}{2}
           \tr{(\phi^\dagger \tilde{\phi}+h.c.)}+\tilde{M}_3^2
           \sigma^\dagger\sigma \nonumber \\
& & +\lambda_1[(\chi_L^\dagger \chi_L)^2+(\chi_R^\dagger \chi_R)^2]
    +2\lambda_2 (\chi_L^\dagger \chi_L)(\chi_R^\dagger \chi_R) \nonumber \\
& & +\frac{1}{2}\lambda_3[\chi_L^\dagger(Y_3\phi+Y_4 \tilde{\phi})\chi_R
       \sigma^\dagger + h.c.] \nonumber \\
& &  +\lambda_4 \large[\chi_L^\dagger(Y_3\phi+Y_4 \tilde{\phi})
           (Y_3\phi^\dagger+Y_4\tilde{\phi}^\dagger)\chi_L \nonumber \\
& &  \hspace{4ex}  +\chi_R^\dagger(Y_3\phi^\dagger+Y_4\tilde{\phi}^\dagger)
                  (Y_3\phi+Y_4 \tilde{\phi})\chi_R \large] \nonumber \\
& & +\lambda_5(\chi_L^\dagger \chi_L+\chi_R^\dagger \chi_R)\tr(\phi^\dagger
   \phi)
    +\lambda_6(\chi_L^\dagger\chi_L+\chi_R^\dagger \chi_R)\sigma^\dagger
      \sigma\nonumber \\
& & +\lambda_7' \tr(\phi^\dagger\phi\phi^\dagger\phi)
    +\frac{1}{3}\lambda_8'\tr(\phi^\dagger\tilde{\phi}
      \tilde{\phi}^\dagger\phi)
    +\frac{1}{12}\lambda_8'
        [\tr(\phi^\dagger\tilde{\phi}\phi^\dagger\tilde{\phi})+h.c.]
  \nonumber \\
& &  +\frac{1}{2} \lambda_9
        [\tr(\phi^\dagger\phi\phi^\dagger\tilde{\phi})+h.c.]
    +\lambda_0
        [\tr(\phi^\dagger\phi)]^2 + \lambda_{10}(\sigma^\dagger\sigma)^2~,
\label{Veff}
\eea
with the mass term corrections and quartic couplings $\lambda_i$ given
as functions of Yukawa couplings and the cutoff.
For our purposes we need only the following expressions for the potential
parameters:
\bea
\tilde{M}_0^2 & = & M_0^2-\frac{1}{8\pi^2}\left [Y_5^2-\frac{3}{8}
Z_{\chi}(3g_2^2+g_1^2)\right ] (\Lambda^2-\mu^2) ~,
\label{M02} \\
\tilde{M}_1^2 & = & M_0^2-\frac{1}{8\pi^2}\left \{ \left [N_c(Y_1^2+Y_2^2)+
(Y_3^2+Y_4^2)\right ]-\frac{9}{4}Z_{\phi}g_2^2\right \}(\Lambda^2-\mu^2)~,
  \nonumber \\  &&
\label{M12}  \\
\tilde{M}_2^2 & = & M_2^2-\frac{1}{4\pi^2}(N_c Y_1 Y_2+Y_3 Y_4)
(\Lambda^2-\mu^2)~,
\label{M22} \\
\tilde{M}_3^2 & = & M_0^2-\frac{1}{4\pi^2}\,Y_6^2 \, (\Lambda^2-\mu^2)~,
\label{M32} \\
\lambda_1 & = & \sixtpi\[ Y_5^4-\frac{3}{16}(3g_2^4+2g_2^2g_1^2+
g_1^4)Z_\chi^2\]
\lnlambda ~, \nonumber \\
\lambda_2 & = & \sixtpi\[ Y_5^4-\frac{3}{16}g_1^4 Z_\chi^2 \] \lnlambda
 \nonumber ~, \\
\lambda_0 & = & \sixtpi\[-\frac{3}{2}g_2^4 Z_\phi^2\] \lnlambda 
 \nonumber ~, \\
\lambda_5 & = & \sixtpi\[ -\frac{9}{8} g_2^4 Z_\phi Z_\chi \] \lnlambda ~,
  \nonumber \\
\lambda_7' & = &\sixtpi\[ N_c (Y_1^4+Y_2^4)+(Y_3^4+Y_4^4)\]\lnlambda ~,
 \nonumber  \\
\lambda_7 & = & \lambda_7' + \lambda_0~.
\label{Lambda}
\eea
Here $g_2$ and $g_1$ are the SU(2) and $\mbox{U(1)}_{B-L}$ gauge couplings,
respectively. The scalar wave-function renormalization constants 
are given by
\bea
Z_{\phi} & = & \sixtpi\, \[N_c (Y_{1}^{2}+Y_{2}^{2})+Y_{3}^{2}+Y_{4}^{2}\]\,
\lnlambda\,, \nonumber \\
Z_{\chi} & = & \sixtpi\, Y_{5}^{2}\, \lnlambda \, , \quad
Z_{\sigma}\; =\;  \sixtpi\, 2 Y_6^2\, \lnlambda~.
\label{zfactors}
\eea
The above parameters are running parameters, depending on the energy scale 
$\mu$, and parametrize the effective Lagrangian at that energy 
scale~\footnote{This bubble-approximation running exactly coincides with
the running one would get from 1-loop \RG equations keeping only trace terms
in the relevant $\beta$ functions and imposing the compositeness boundary
conditions~\cite{BHL}.}.
At $\mu \to \Lambda$ the kinetic terms and quartic couplings of the scalar
fields vanish, their mass terms are driven towards their bare values, and one
recovers the Lagrangian with auxiliary static scalar fields.

Spontaneous symmetry breaking generally occurs when scalar mass terms 
become negative, in the presence of quartic scalar interactions.
While the bare mass parameters $M_i^2$ in equation~(\ref{Laux}) are 
positive, the corresponding running quantities $\tilde{M}_i^2$, given 
by equations~(\ref{M02}) -- (\ref{M32}), may become negative at low energy 
scales provided that the corresponding Yukawa couplings are large enough. 
Those values for which this occurs at $\mu=0$ we shall call the 
{\em critical } Yukawa couplings. Of course, in 4F-language this correspond 
to the critical values of the original 4F-couplings. For
$\tilde{M}_i^2$ to become negative at some scale $\mu^2 > 0$ the corresponding
Yukawa couplings, or combinations of them, must be above their critical values.
If this is to happen at scales $\mu \ll \Lambda$ the Yukawa couplings
must be fine-tuned very closely to their critical values to ensure
the proper cancellation between the large bare masses of the scalars
and the $\Lambda^2$ corrections in equations~(\ref{M02}) -- (\ref{M32}).
This is equivalent to the usual fine-tuning problem of gauge theories with
elementary Higgs scalars~\cite{BHL}.

We assume that the scale $\mu_R$ at which parity gets spontaneously broken
(i.e. $\chi_R^0$  develops a vacuum expectation value) is higher than the \EW 
scale $\mu_{EW}\sim 100$~GeV, i.e. that $\tilde{M}_0^2$ changes its sign at
a higher scale than $\tilde{M}_1^2$. This means that
$Y_5^2 - (3/8)Z_\chi(3g_2^2+g_1^2)$ should be bigger than
$\tilde{Y}^2-\frac{9}{4}Z_{\phi}g_2^2$ [see equations~(\ref{M02}) and 
(\ref{M12})], where $\tilde{Y}^2 \equiv N_c(Y_1^2+Y_2^2)+(Y_3^2+Y_4^2)\;$.
By analysing the minima of the effective potential (\ref{Veff}) one can
show that if the condition
\beq
Y_5^2-\frac{3}{8}Z_\chi(3g_2^2+g_1^2)>2\,Y_6^2~
\label{condit}
\eeq
is satisfied, either $\chi_R$ or $\chi_L$ (but not both) acquire a vacuum
expectation value but the $\sigma$ field does not, whereas for the opposite 
condition $\sigma$ acquires a non-zero vacuum expectation value, but not 
$\chi_R$ or $\chi_L$. Clearly the latter situation is phenomenologically 
unacceptable, but by choosing the four-fermion couplings $G_7$ and $G_8$ 
accordingly we can easily satisfy equation~(\ref{condit}).

\subsubsection{The Case $\kappa' =0$}
\label{sec:knull}

Let us now discuss the vacuum structure below the \EW breaking scale.
The non-vanishing vacuum expectation values are 
$v_R$, $\kappa$ and $\kappa'$. Since $m_t \gg
m_b$, it follows from  equation~(\ref{fmass}) that $\kappa$ should be much 
larger than $\kappa'$ or vice versa provided no significant cancellation 
between $Y_1\kappa'$ and $Y_2\kappa$ occurs. Without loss of generality 
one can take $\kappa > \kappa'$. To further simplify the discussion, we 
will first consider the frequently used assumption~\cite{LR} $\kappa'=0$. 
The relation $m_t \gg m_b$ then translates into $Y_1 \gg Y_2$. In the 
conventional approach this assumption does not lead to any contradiction 
with phenomenology. However, as we shall see, in the composite model the 
condition $\kappa'=0$ cannot be exact.

Consistency of the first-derivative conditions with $\kappa'=0$ 
requires $Y_1 Y_2=0$, $Y_3 Y_4=0$ and $M_2^2=0$ (this gives 
$\tilde{M}_2^2=\lambda_9=0$, and as follows from equation~(\ref{Veff}), 
all the terms in the effective potential which are linear in $\kappa'$ 
become zero in this limit, as they should). The condition $Y_1 Y_2=0$ 
along with $\kappa'=0$ implies that either $Y_1=0$, $m_t=0$ or $Y_2=0$, 
$m_b=0$. The first possibility is obviously phenomenologically 
unacceptable, whereas the second one can be considered as a reasonable 
first approximation. Therefore we assume $Y_1 \neq 0$ and $Y_2=0$. 
The situation is less clear for the lepton Yukawa couplings $Y_3$ 
and $Y_4$. Since $m_\tau \ll m_t$ and the Dirac mass $m_D$ of 
$\nu_\tau$ is unknown, one can choose either $Y_3 \neq 0$, $Y_4=0$ 
or $Y_3=0$, $Y_4 \neq 0$. It turns out that the vacuum stability 
condition in this model requires $Y_4^2 >Y_3^2$, therefore we choose 
$Y_3=0$ and $Y_4\neq 0$.

For $\sigma_0=v_L=\kappa'=Y_2=Y_3=0$ one can easily find expressions for 
the vacuum expectation values 
of $\chi_R$ and $\phi$. Approximate expressions in terms of the 
parity breaking scale $\mu_R$ and the \EW breaking scale $\mu_{EW}$ are
\beq
v_R^2 \simeq \left(\frac{M_0^2}{\Lambda^2}\right)
\frac{\mu_{R}^2}{2\lambda_1}\,, \quad\quad
\kappa^2\simeq \left( \frac{M_0^2}{\Lambda^2}\right)
\frac{\mu_{EW}^2}{2\lambda_7}~,
\label{k20}
\eeq
and the ratio of the squared vacuum expectation values can be written as
\beq
\frac{\kappa^2}{v_R^2}\simeq \left(\frac{\lambda_1}{\lambda_7}\right)
\frac{\mu_{EW}^2}{\mu_R^2}\sim \frac{\mu_{EW}^2}{\mu_R^2}\simeq
\frac{|\lambda_5|}{2\lambda_1}+\frac{\mu_1^2}{\mu_R^2}~.
\label{ratio}
\eeq

The parity breaking scale $\mu_R$ is the scale where
the effective mass term $\tilde{M}_0^2$ becomes negative for a given
Yukawa coupling $Y_5>(Y_5)_{crit}$ (formally $\mu_R^2<0$ for sub-critical
$Y_5$), while $\mu_1$ is the scale, different from $\mu_{EW}$, where this
happens for the mass term $\tilde{M}_1^2$ and a given $\tilde{Y}^2$.

Recall now that in conventional \LR models with $\mu_{EW} \ll \mu_R \ll
\Lambda_{GUT}$ (or $\Lambda_{\rm Planck}$) one has to fine-tune two
gauge hierarchies: $\Lambda_{GUT}\dash \mu_R$ and $\mu_R\dash \mu_{EW}$.
The situation here is quite similar: to achieve $\mu_{EW} \ll \mu_R
\ll \Lambda$ one has to fine-tune two Yukawa couplings, $Y_5^2$ and
$\tilde{Y}^2$. Tuning of $Y_5^2$ allows for the hierarchy
$\mu_R^2 \ll \Lambda^2$; one then needs to adjust
$\tilde{Y}^2$ (or $\mu_1^2$) to achieve $\mu_{EW}^2 \ll \mu_R^2$ through
equation~(\ref{ratio}).

Since $\lambda_5$ only contains relatively small gauge couplings while
$Y_5 \sim \cal{O}$$(1)$, we typically have $|\lambda_5|/(2\lambda_1)
\sim 10^{-2}$. Thus, if there is no significant cancellation between the
two terms in (\ref{ratio}), one obtains a right-handed scale of the order 
of a few TeV. Unfortunately, such a low \LR scale scenario is not viable.
As we shall see below, the squared masses of two Higgs bosons become 
negative (i.e. the vacuum becomes unstable) unless $v_R\,\gtap \,20$~TeV. 
This requires some cancellation~\footnote{Notice that this does not 
increase the number of the parameters to be tuned but just shifts the 
value to which one of them should be adjusted.} in equation~(\ref{ratio}), 
and then the right-handed scale $v_R\sim \mu_R$ can in
principle lie anywhere between a few tens of TeV and $\Lambda$.
However, if one prefers ``minimal cancellation'' in equation~(\ref{ratio}),
by about two orders of magnitude or so, one would arrive at a value of
$v_R$ around $20$~TeV. In any case it is interesting that the partial
cancellation of the two terms in (\ref{ratio}) implies $\mu_1^2<0$,
i.e. that $\tilde{Y}^2$ must be below its critical value. This means that
$\tilde{M}_1^2$ never becomes negative. In fact it is the $\tilde{M}_0^2$
term, responsible for parity breakdown, that also drives the vacuum 
expectation value of the
bi-doublet. It follows from the condition $\partial V_{\rm eff}/\partial
\kappa=0$ that the effective driving term for $\kappa$ is $\tilde{M}_1^2+
\lambda_5 v_R^2$; it may become negative for large enough
$v_R^2$ even if $\tilde{M}_1^2$ is positive (remember that $\lambda_5<0$).
Thus one finds a tumbling scenario where the breakdown of parity and 
$\mbox{SU(2)}_R$ occurring at the scale $\mu_R$ causes the breakdown of 
the \EW symmetry at a lower scale $\mu_{EW}$.

\subsubsection{The Case $\kappa' \ne 0$}

After having established the vacuum structure for the simplified case
$v_R, \kappa \ne 0$, $\kappa'=0$, we now turn to the more general 
case with $\kappa' \ne 0$. Under the assumption that the righthanded
scale is sufficiently above the \EW scale, the details of parity 
breakdown should remain unchanged. Analysing the minimum conditions 
in the $\kappa-\kappa'$-sector, one finds that the ratio of 
vacuum expectation values is
approximately given by the ratio of lepton Yukawa couplings:
\beq
\frac{\kappa'}{\kappa}=
-\frac{Y_3}{Y_4} + {\cal O} \(\frac{\kappa^2}{v_R^2}\)~.
\label{y3y4bez}
\eeq
Provided the condition~(\ref{condit}) is fulfilled, one can 
show~\cite{ALSV2} that $v_L$ and $\sigma_0$ also remain zero after \EW 
symmetry breaking. The diagonalization of the scalar mass matrices 
leads to positive values for the squared masses, except zero for the
expected \GBsk and thus the described situation is 
a true minimum of the model. 

\subsection{Fermion Masses}

Now we turn to the calculation of fermion and scalar masses, first
in the simplified limit. To calculate physical observables one should 
first rescale the Higgs fields by absorbing the $Z$-factors in 
equation~(\ref{Leff}) into the definitions of the scalar fields to 
bring their kinetic terms into the canonical form. This amounts to 
dividing the (mass)$^2$ terms by the corresponding $Z$-factors, 
Yukawa couplings by $\sqrt{Z}$ and multiplying the scalar fields
and their vacuum expectation values 
by $\sqrt{Z}$. Renormalization factors of the quartic
couplings depend on the scalars involved and can be readily read off 
from the effective potential. We will use hats $(\,\hat{ }\,)$ to denote
quantities in the new normalization.

As pointed out earlier, the minimization of the effective Higgs 
potential gives $\sigma_0=0=v_L$. This means that the entries 
$\beta$ and $\tilde{\mu}$ in the neutrino mass matrix (\ref{numass1}) 
are zero. As a result one finds an exactly massless neutrino eigenstate 
and two heavy Majorana neutrinos with degenerate masses 
$\sqrt{M_{\nu_R}^{2}+m_{D}^{2}}$ and opposite CP-parities which combine to 
form a heavy Dirac neutrino. Since $m_D \ll M_{\nu_R}$ the \EW eigenstate 
$\nu_L\equiv \nu_{\tau}$ is predominantly the massless eigenstate,
whereas the right-handed neutrino $\nu_R$ and the singlet fermion $S_L$
consists predominantly of the heavy eigenstates. 

As mentioned before, for $\kappa'=0$ we have $Y_2=0=Y_3$. This yields 
$m_t=Y_1\kappa$, $m_{\tau}=Y_4\kappa$ and $m_b=m_D=0$. Vanishing Dirac 
neutrino mass $m_D$ implies the absence of neutrino mixing, and the heavy 
neutrino mass is now $M_{\nu_R}=Y_{5}v_{R}$. From
equations~(\ref{fmass}) and (\ref{zfactors}) and the definition of the
renormalized Yukawa couplings one can readily find
\bea
\hat{\kappa}^2 &=& (174~{\rm GeV})^{2}    \nonumber \\
 &=& N_c m_{t}^{2}\(1+\frac{Y_{4}^{2}}{N_{c}Y_{1}^{2}}\)\sixtpi\lnlambda 
\nonumber \\
& \approx & \frac{m_{t}^{2}N_c}{16\pi^2}\lnlambda \equiv m_{t}^{2} N_c l_0~.
\label{mtopbubble}
\eea
Here $\hat{\kappa}$ (or $\sqrt{\hat{\kappa}^2+\hat{\kappa}'^2}$ for
$\kappa'\neq 0$) is identified with the usual \EW vacuum expectation
value and the top quark mass is fixed in terms of this known vacuum 
expectation value and the scale of new
physics $\Lambda$. Note that this expression coincides
with the one derived in bubble approximation by BHL~\cite{BHL}. From 
equation~(\ref{mtopbubble}) it follows that the top quark mass depends on
$\Lambda$ logarithmically. For example, for $\Lambda=10^{15}$~GeV one finds
$m_t\simeq 165$~GeV. However, as in the BHL-model, the \RG analysis will 
change this result considerably.

Using considerations similar to those which led to 
equation~(\ref{mtopbubble}), it is easy to obtain the following relation 
between the right-handed vacuum expectation value $v_R$, the heavy neutrino 
mass $M_{\nu_R}$ and the scale $\Lambda$:
\beq
\hvr2 =  M_{\nu_R}^{2}~\sixtpi\lnlambda~ = M_{\nu_R}^2 \cdot l_0~.
\label{Vr2}
\eeq
Note that $\mu \approx m_t$ is understood in equation~(\ref{mtopbubble}),
whereas $\mu \approx M_{\nu_R}$ in equation~(\ref{Vr2}). We will assume
$m_t,M_{\nu_R}\ll \Lambda$ and $M_{\nu_R}/m_t \ll \Lambda/M_{\nu_R}$  
throughout this article,
therefore $\ln\frac{\Lambda^2}{m_t^2} \approx \ln\frac{\Lambda^2}
{M_{\nu_R}^2}$, i.e. the logarithmic factor $l_0$ is universal. Then from 
equations~(\ref{mtopbubble}) and (\ref{Vr2}) one finds
\beq
\frac{{\hvr2}}{M_{\nu_R}^2}\approx 
\frac{1}{3}\frac{\hat{\kappa}^2}{m_{t}^2}~.
\label{VrtoM}
\eeq
Notice that the central top mass value from the Fermilab 
collaborations~\cite{CDFD0}, $m_t$=175~GeV, implies $l_0 \approx 1/3$. 
The mass of the $\tau$ lepton is not predicted in this model since it is
only weakly coupled to the bi-doublet; it is given by $m_{\tau} = (Y_4/Y_1)
m_t$ and can be adjusted to a desirable value by choosing the proper
magnitude of the ratio $Y_4/Y_1$, or $G_3/G_1$.

The composite Higgs bosons in this model include the would-be \GBs
$G_1^\pm \approx \chi_R^\pm$ (eaten by $W_R^\pm$), $G_2^\pm = \phi_1^\pm$
(eaten by $W_L^\pm$), $G_1^0 = \chi_{Ri}^0$ (eaten by $Z_R$) and  $G_2^0 =
\phi_{1i}^0$ (eaten by $Z_L$). The physical Higgs boson sector of the model
contains two CP-even neutral scalars $H_1^0 \approx \chi_{Rr}^0$ and
$H_2^0 \approx \phi_{1r}^0$ with the masses
\bea
M_{H_1^0}^2 & \simeq & 4M_{\nu_R}^{2}\[1-\frac{3}{16}
\(3g^{4}+2g^{2}g'^{2}+g'^{4}\)l_0^2\]\approx 4M_{\nu_R}^2 \;,\label{MH10}\\
M_{H_2^0}^2 & \simeq & 4 m_t^2 \(1-\frac{m_{\tau}^{2}}{3 m_{t}^{2}} -
\frac{9}{4}g^4 l_0^2 \) \approx 4 m_{t}^{2}\;, \label{MH20}
\eea
which are directly related to the two steps of symmetry breaking,
$\mbox{SU(2)}_R\times \mbox{U(1)}_{B-L}\rightarrow \mbox{U(1)}_Y$ and
$\mbox{SU(2)}_L\times \mbox{U(1)}_Y    \rightarrow \mbox{U(1)}_{em}$. 
The mass of the scalar $H_2^0$, which is the analog of the \SM Higgs boson
[equation~(\ref{MH20})], essentially coincides with the one obtained in the
bubble approximation by BHL~\cite{BHL}. This simply reflects the fact that
this boson is the $t\bar{t}$ bound-state with a mass of  $\approx 2 m_t$.
Analogously, the mass of the heavy CP-even scalar $H_1^0 \approx
\chi_{Rr}^0$ is approximately $2M_{\nu_R}$ since it is a bound-state of 
heavy neutrinos.

Further, there are the charged Higgs bosons $H_3^\pm \approx \phi_2^\pm$ 
with their neutral CP-even and CP-odd partners 
$H_{3r}^0 =\phi_{2r}^0$ and $H_{3i}^0 =\phi_{2i}^0\;$, and finally the 
$\chi_L$-fields $H_4^\pm = \chi_L^\pm\;$, $H_{4r}^0=\chi_{Lr}^0$  and
$H_{4i}^0=\chi_{Li}^0$ with the masses
\bea
M_{H_3^\pm}^2 &\approx&  \frac{2}{3}M_{\nu_R}^{2}~
      \frac{m_{\tau}^{2}}{m_{t}^{2}}\;,
\label{MH3pm} \\
M_{H_{3r}^0}^2 &=& M_{H_{3i}^0}^2 \approx \frac{2}{3}M_{\nu_R}^{2}~
\frac{m_{\tau}^{2}}{m_{t}^{2}} - \frac{1}{2} M_{H_2^0}^2~\;,\label{MH30}\\
M_{H_4^\pm}^2 &=& \frac{3}{8}\(3g_2^4+2g_2^2g_1^2\)l_0^2\,M_{\nu_R}^2
                   +2 m_{\tau}^2
\label{MH4pm} \\
M_{H_{4r}^0}^2&=& M_{H_{4i}^0}^2 =
\frac{3}{8}\(3g_2^4+2g_2^2g_1^2\)l_0^2\,M_{\nu_R}^2~. \label{MH40}
\eea
In conventional \LR models only one scalar, which is the analog of the 
\SM Higgs boson, is light (at the \EW scale), all the others have their 
masses of the order of the right-handed scale 
$M_{\nu_R}$~\cite{LR1,LR3,LR,LRdesh}. In this composite case, the masses 
of those scalars are also proportional to $M_{\nu_R}$, but all of them 
except the mass of $H_1^0$ have some suppression factors. The mass of
the charged scalars $H_3^\pm \approx \phi_2^\pm$ is suppressed by the factor
$m_{\tau}/m_t$ and is therefore of the order $10^{-2} M_{\nu_R}$. 
The masses of the neutral $H_{3r}^0$ and $H_{3i}^0$ are even smaller; 
they are related to the masses of the charged $H_3^\pm$ and the \SM Higgs 
$H_2^0$ by equation~(\ref{MH30}). From the vacuum stability condition 
$M_{H_{3}^0}^2>0$ one thus obtains an upper limit on the \SM Higgs boson 
mass $M_{H_2^0}$ (for a given $M_{\nu_R}$) or a lower limit on the 
right-handed mass $M_{\nu_R}$ (for a given $M_{H_2^0}$). For example, for 
$M_{H_2^0} \approx 60$~GeV one finds $M_{\nu_R}\gtap 5$~TeV. 
However, since in the top condensate approach the mass of
the \SM Higgs is around $2m_t$ (or $\sim m_t$ after the \RG improvement), 
one obtains a stronger lower bound on the right handed gauge symmetry 
breaking scale $M_{\nu_R}$ of about $20-50$~TeV.

The masses of the $\chi_L$ scalars [equations~(\ref{MH4pm}),(\ref{MH40})]
vanish in the limit \mbox{$(\lambda_2-\lambda_1) \to 0$} (i.e. $g_2 \to 0$) 
and $m_\tau \to 0$. This fact has a simple interpretation. In the limit
$\lambda_2=\lambda_1$  (which corresponds to the fermion-bubble level)
the ($\chi_L,\chi_R$) sector of the effective Higgs potential
[equation~(\ref{Veff})] depends on $\chi_L$ and $\chi_R$ only through the
combination $(\chi_L^\dagger \chi_L+\chi_R^\dagger \chi_R)$. This implies
that the potential has a global SU(4) symmetry which is larger than the
initial $\mbox{SU(2)}_L\times \mbox{SU(2)}_R\times \mbox{U(1)}_{B-L}$ 
symmetry. After $\chi_R^0$ gets a non-vanishing vacuum expectation value 
$v_R$, the symmetry 
is broken down to SU(3), resulting in $15 - 8=7$ \GBsp Three of them 
($\chi_R^\pm$ and $\Im \chi_R^0$) are eaten by the $\mbox{SU(2)}_R$ gauge 
bosons $W_R^\pm$ and $Z_R$, and the remaining four 
($\chi_L^\pm$, $\Re \chi_L^0$ and $\Im \chi_R^0$) are physical massless 
\GBsp The SU(4) symmetry is broken by the $\phi$--dependent
terms in the effective potential and by SU(2) gauge interactions.
As a result, $\chi_L^\pm$, $\Re \chi_L^0$ and $\Im \chi_L^0$ acquire small
masses and become pseudo-\GBsp In fact, the origin of this approximate
SU(4) symmetry can be traced back to the four-fermion operators of 
equation~(\ref{L4f}). It is an accidental symmetry resulting from the 
gauge invariance and parity symmetry of the $G_7$ term. Note that no such 
symmetry occurs in conventional \LR models.

After relaxing the approximation $\kappa'=0$, one obtains non-vanishing
masses $m_b$ and $m_D$ (notice that the Yukawa couplings $Y_2$ and $Y_3$
will also be non-zero in this case). However, these masses are not predicted
in the model and can simply be adjusted to desirable values. The Dirac
neutrino mass $m_D$ is unknown and so remains a free parameter; however,
it must be smaller than $m_{\tau}$ in this model in order to satisfy a
vacuum stability condition $Y_4^2-Y_3^2>0$, which one finds
from analysing the potential for the general case~\cite{ALSV2}.
This condition is equivalent to $m_{\tau}^2-m_D^2>0$. Also, one finds that 
the lower bound on $M_{\nu_R}$ is strengthened for $\kappa' \ne 0$.
The Higgs boson masses and mass eigenstates for the general case 
$\kappa' \ne 0$ are only slightly modified and can be also be 
found in a recent publication~\cite{ALSV2}. 

\subsection{Renormalization Group Analysis}
\label{sec:fixp}

So far the model was analysed in the ``bubble approximation'', where only 
fermion and $\mbox{SU(2)}_L\times \mbox{SU(2)}_R \times \mbox{U(1)}_{B-L}$ 
gauge boson loops contribute. However, important corrections arise from 
QCD effects and loops with composite Higgs scalars. Those effects can be 
accounted for~\cite{BHL} by solving the full one-loop \RG 
equations of the low energy effective theory with boundary conditions 
corresponding to compositeness.

These boundary conditions follow from the vanishing of the radiatively 
induced kinetic terms for the Higgs scalars at the scale $\Lambda$, 
where the composite particles break up into their constituents:
\begin{equation}
Z_{\phi}(\mu^2 \to \Lambda^2)=Z_{\chi}(\mu^2 \to \Lambda^2)=
Z_{\sigma}(\mu^2 \to \Lambda^2)=0~.
\label{zboundary}
\end{equation}
After rescaling the scalar fields to bring their kinetic terms
into the canonical form (i.e. normalized to one) one obtains corresponding 
boundary conditions for the Yukawa and quartic couplings of the low energy 
effective Lagrangian. These conditions are similar to those 
obtained by BHL and have the following generic form:
\begin{equation}
\hat{Y}^2 = \frac{Y^2}{Z} \stackrel{\mu^2\rightarrow\Lambda^2}
{\longrightarrow}\infty~,\;\;
\hat{\lambda}=\frac{\lambda}{Z^2}\stackrel{\mu^2\rightarrow\Lambda^2}
{\longrightarrow} \infty~,\;\;
\frac{\hat{\lambda}}{\hat{Y}^4}=\frac{\lambda}{Y^4}
\stackrel{\mu^2\rightarrow\Lambda^2}{\longrightarrow} 0~.
\label{comp3}
\end{equation}
The renormalized parameters of the model derived in bubble 
approximation already satisfy the compositeness conditions; for 
example, the renormalized Yukawa couplings are
\begin{eqnarray}
\hat{Y}_1^2(\mu) & = & \frac{Y_1^2}{Z_\phi}
        \approx \left[\frac{3}{16\pi^2} \ln \left( 
        \frac{\Lambda^{2}}{\mu^{2}} \right) \right]^{-1}~;
                \;\;\;\;(\hat{Y}_4 \ll \hat{Y}_1)~,
\label{boundcon1} \\
\hat{Y}_4^2(\mu) & \approx & \frac{Y_4^2}{Y_1^2}\; \hat{Y}_1^2(\mu)
                         =   \frac{G_3}{G_1}    \; \hat{Y}_1^2(\mu)~,
\label{boundcon2}  \\
\hat{Y}_5^2(\mu) & = & \frac{Y_5^2}{Z_\chi} =\left[
         \frac{1}{16\pi^2} \ln \left( 
         \frac{\Lambda^{2}}{\mu^{2}} \right) \right]^{-1}~,
\label{boundcon3} \\
\hat{Y}_6^2(\mu) & = & \frac{Y_6^2}{Z_\sigma} =\left[
             \frac{2}{16\pi^2} \ln \left( 
             \frac{\Lambda^{2}}{\mu^{2}} \right) \right]^{-1}
\label{boundcon6}~.
\end{eqnarray}
Obviously they diverge as $\mu\rightarrow\Lambda$. Furthermore, their
running coincides exactly with that described by the fermion loop
contributions (the trace terms) to one-loop $\beta$-functions of the 
corresponding \LRS theory. The idea is now to identify the Landau poles in 
the {\em full} one-loop \RG evolution of couplings with the compositeness 
scale $\Lambda$ and run the couplings down to low energy scales.

We will first consider the simplified scenario with $\kappa'=0$. In the 
one-generation scenario the \RG equations for the 
Yukawa couplings in the limit $\kappa'=0$ (which requires $Y_2=Y_3=0$, 
see section~\ref{sec:knull}) reduce to~\footnote{In the following we
will omit the hats over the renormalized quantities.}
\renewcommand{\thefootnote}{\arabic{footnote}}
\begin{eqnarray}
16\pi^{2} \frac{dY_1}{dt} & = &
5Y_1^3+Y_1 Y_4^2 
- \left(8 g_3^2+\frac{9}{2}g_2^2+\frac{1}{6}g_1^2\right)Y_1~,
\label{betatrun1}\\
16\pi^{2} \frac{dY_4}{dt} & = &
3Y_4^3+3 Y_4 Y_1^2+Y_4 Y_5^2 -
\left(\frac{9}{2}g_2^2+\frac{3}{2}g_1^2\right)Y_4~,
\label{betatrun2}\\
16\pi^{2} \frac{dY_5}{dt} & = &
 \frac{7}{2}Y_5^3 + Y_5\left(Y_4^2+[2Y_6^2]\right)
              -\left(\frac{9}{4}g_2^2+\frac{3}{4}g_1^2\right)Y_5~,
\label{betatrun3}\\
16\pi^{2} \frac{dY_6}{dt} & = &
 [6]Y_6^3 + 4Y_6 Y_5^2~.
\label{betatrun4}
\end{eqnarray}
For large values of $Y_i$ the $Y_i^3$--contributions in the 
$\beta$-functions are dominant; they quickly drive the couplings 
down to values of order one as the scale $\mu$ decreases. In this 
regime gauge and other Yukawa coupling contributions become 
important, and the interplay of these contributions and $Y_i^3$ 
terms result in so-called infrared quasi-fixed 
points~\cite{quasiIR}. Thus a large range of initial values of 
the Yukawa couplings at the cutoff is focused into a small range 
at low energies. The masses of the fermions will then be given 
implicitly by conditions of the kind $Y(m)\cdot \mbox{VEV} = m$.

To evolve the Yukawa couplings with one-loop $\beta$-functions to 
their Landau poles, i.e. in the non-perturbative regime, may 
appear questionable. However, it has been 
argued~\cite{BHL,manus} that this should not result in any significant 
errors. First the running time $t=\ln\mu$ in the non-perturbative 
domain is only a few percent of the total running time. Second, 
and more importantly, the infrared quasi-fixed point structure of 
the \RG equations makes the predictions fairly insensitive to the 
detailed behaviour of the solutions in the large Yukawa coupling 
domain. Lattice gauge theory has generally confirmed the 
reliability of perturbation theory in this fixed point 
analysis~\cite{latticeFP}.

Except for switching to the \SM \RG equations below 
the parity breaking scale the only relevant threshold effects in the 
evolution are due to the masses of the $\sigma$ scalars. From the 
vacuum structure analysis in bubble approximation, the vacuum
expectation values of 
$\sigma$ and $\chi_R$ do not coexist; phenomenology then dictates the 
choice $\sigma_0=0$, $v_R \neq 0$, which requires the four-fermion 
coupling $G_8$ to be sub-critical, or at least satisfying the condition 
(\ref{condit}). We will assume that the same holds true beyond the bubble 
approximation and consider $\sigma$ to be non-propagating, or at least 
decoupled from the low-energy spectrum of the model. Therefore the 
effects from propagating $\sigma$ scalars can be switched off directly 
at the cutoff by neglecting the contributions in square brackets of 
equations~(\ref{betatrun3}) and (\ref{betatrun4}). In this limit the 
running of $Y_6$ does not influence the running of the other 
Yukawa couplings.

In numerical calculations a large number $Y_i(\Lambda)$ must be 
used as boundary condition for Yukawa couplings instead of infinity. 
Fortunately, the infrared quasi-fixed point structure of the 
\RG equation makes the solutions fairly insensitive to 
the actual values of $Y_i(\Lambda)$ provided that they are large 
enough~\cite{BHL,quasiIR}. In fact, the infrared quasi-fixed-point 
behaviour sets in already for 
$Y_i(\Lambda)\raisebox{-.4ex}{\rlap{$\sim$}}\raisebox{.4ex}{$>$} 5$. 
The calculations were performed using $Y_5(\Lambda)=10$, but taking e.g. 
$10^3$ instead of 10 results only in a correction of about $0.4\%$ 
in the low-energy value of $Y_5$. The fermion-loop results of
equations~(\ref{boundcon1})--(\ref{boundcon6}) imply a fixed ratio 
between the running coupling constants which one could, as a first 
approximation, also impose as the boundary condition at the cutoff 
for the full \RG evolution, e.g. 
$Y_1(\Lambda)=3\,Y_5(\Lambda)=30$. Fortunately, once again, the 
numerical results depend very weakly on this scaling factor, and for 
this purpose it could just as well be taken to be unity.
\begin{figure}[htb]
\centerline{
\epsfxsize=60ex
\rotate[r]
{\epsffile{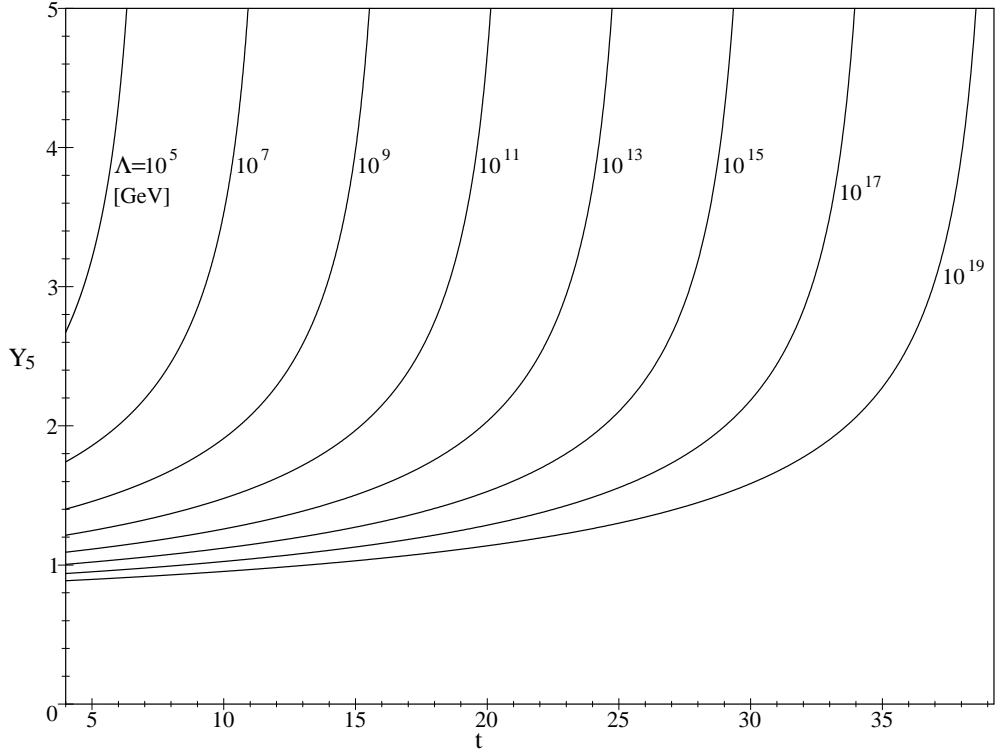}}
}
\caption[]{\small\sl
Renormalization group evolution of the $Y_5$ Yukawa coupling for 
various compositeness scales $\Lambda,\; t=\ln(\mu /m_Z)$.
\label{fig:Y5fixed}}
\end{figure}
Figure~\ref{fig:Y5fixed} shows $Y_5(\mu)$ obtained by numerically 
solving equations~(\ref{betatrun1})--(\ref{betatrun3}) for various 
values of the cutoff $\Lambda$. The heavy neutrino mass $M_{\nu_R}$ is 
determined by the equation
\begin{equation}
M_{\nu_R} = Y_5(M_{\nu_R})\cdot \hat{v}_R~,
\end{equation}
and for $v_R\sim \mu_R \ll \Lambda$ one finds values of
$Y_5(M_{\nu_R})\simeq Y_5(\mu_R)$ roughly between 1 and 2.

The evolution of $Y_1$ and $Y_4$ below the parity breaking scale is 
determined by the usual \SM $\beta$-functions~\cite{oneloopb}. It 
turns out that the numerically most important difference between 
the \LR and \SM $\beta$-functions for $Y_1$ is a contribution of 
$1/2\; Y_1^3$ coming from the self energy diagram with a $\phi_2^+$ 
scalar exchange. Since the mass of $\phi_2^+$ is not of the order 
of the right-handed scale but is suppressed by a factor 
$\approx 10^{-2}$ we switch to the \SM $\beta$-functions two
orders of magnitude below the parity breaking scale $\mu_R$.
\begin{figure}[htb]
\centerline{
\epsfxsize=60ex
\rotate[r]{ \epsffile{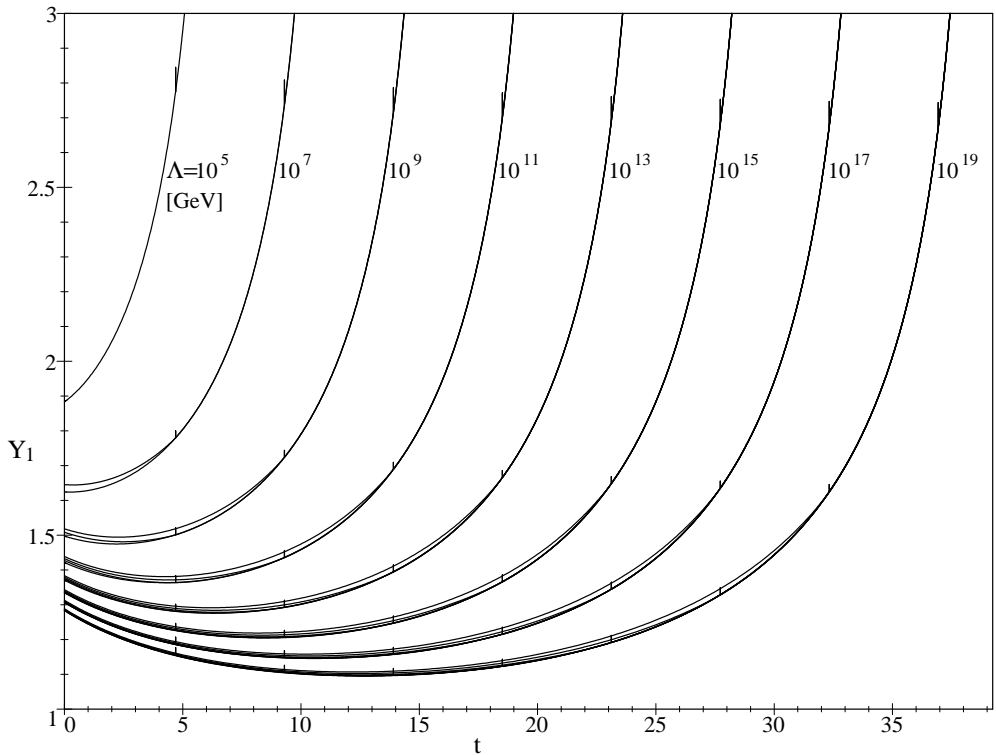} }}
\caption[]{\small\sl
Renormalization Group evolution of the $Y_1$ Yukawa coupling for 
various parity breaking scales $\mu_R$ (indicated by little ticks) 
and compositeness scales $\Lambda;\; t=\ln(\mu /m_Z)$.
\label{fig:Y1fixed}}
\end{figure}

In figure~\ref{fig:Y1fixed} we present the numerical solutions for 
$Y_1(\mu)$ for various values of $\Lambda$ and $\mu_R$. One can 
clearly see the infrared quasi-fixed point structure of the 
solutions. The values of $Y_1$ at $t=0$ $(\mu=m_Z)$ are to some 
extent sensitive to the magnitude of the cutoff but fairly 
insensitive to the scale where parity breaks. This is because in 
fact the previously mentioned contribution to the $Y_1$
$\beta$-function from the $\phi_2^+$ exchange makes only a 
relatively small difference between $(9/2)\,Y_1^3$ in the \SM and 
$5\,Y_1^3$ in the \LRS model with a bi-doublet. For the 
same reason the top quark mass prediction in the $\kappa'=0$ case 
is very similar to the one of the BHL model~\cite{BHL}, which is 
too high compared to the recent experimental results~\cite{CDFD0}. 
Even for a rather high cutoff $\Lambda\approx 10^{17}$~GeV one obtains 
a top quark mass about $229$~GeV.

Because the $\tau$ lepton is much lighter than the top quark, 
the $\tau$-Yukawa coupling is not governed by any fixed point; instead 
its low energy value depends essentially on the value at the cutoff.
Consequently, the $\tau$ contributes only very little to the composite 
Higgs bi-doublet, which is driven by large $Y_1$ 
and not by $Y_4$. In other words, the model exhibits a top 
condensate (along with a heavy-neutrino condensate) rather than 
a tau condensate. One can readily obtain a suitable low-energy value
of $Y_4$ by choosing a proper value of $Y_4(\Lambda)$.
Therefore, although $m_\tau$ is not predicted in 
this framework, it can be easily adjusted to the correct value.

So far it was assumed that only one of the two neutral components of 
the bi-doublet $\phi$ acquires a vacuum expectation value 
($\kappa\ne 0$, $\kappa'=0$). 
Apparently, the resulting fixed-point value of the top quark Yukawa 
coupling is in this limit outside the phenomenologically acceptable 
region~\cite{CDFD0}. Moreover, the assumption $\kappa'=0$ implies 
a zero bottom quark mass. Evidently, for this model to be realistic 
with $m_b \ne 0$ one requires either $\kappa' \ne 0 $ or $Y_2\ne 0$. In 
conventional \LR symmetric models these two conditions are unrelated 
and can be satisfied separately, but in this model $Y_2\ne 0$ 
automatically means $\kappa'\ne 0$ and vice versa. In the following 
we show the results from analysing the \RG evolution of the full set 
of Yukawa couplings for the general case, which indeed leads to viable 
top and bottom quark masses for a range of values of $\kappa' \sim \kappa$.

\begin{figure}[htb]
\centerline{
\epsfysize=68ex
{\epsffile{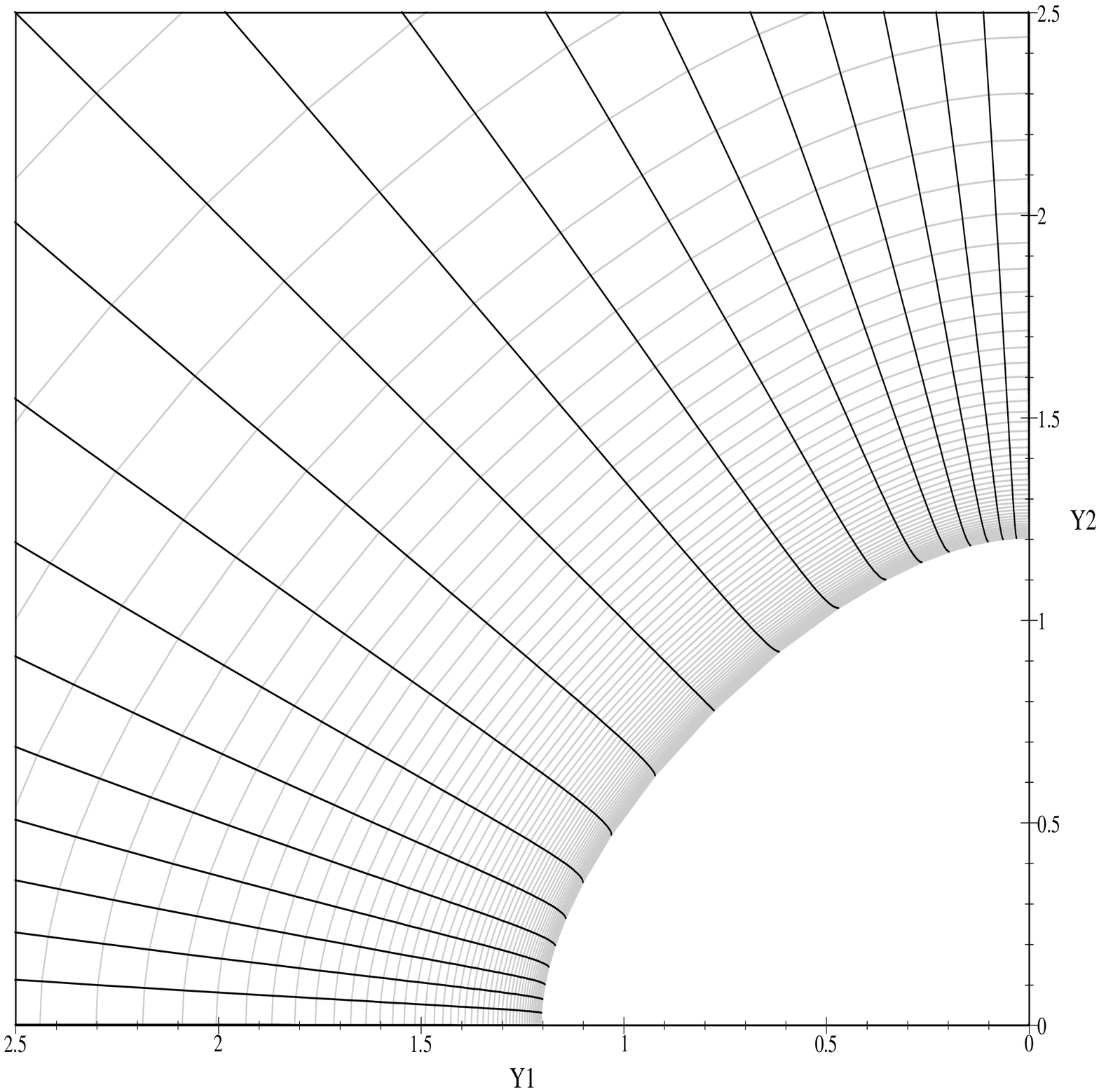}}}
\caption[]{\small\sl
Location of fixed points in the $Y_1$--$Y_2$-plane for 
\mbox{$\Lambda=10^{15}$~{\rm GeV}} and \mbox{$\mu_R=10^6$~{\rm GeV}}.
\label{fig:y1y2plane}}
\end{figure}

As we argued previously, there is a stable vacuum for this model 
with $\kappa, \kappa'$ and $v_R$ non-vanishing. In the limit of 
$\kappa, \kappa' \ll v_R$ the ratio of vacuum expectation values 
is given by 
\begin{equation}
\frac{\kappa'}{\kappa} = - \frac{Y_3}{Y_4}
          + {\cal O}\left(\frac{\kappa^2}{v_{R}^{2}}\right)\;\;.
\label{y34cond}
\end{equation}
Without loss of generality we assume $|\kappa'| < |\kappa|$.
On the other hand, to obtain $m_t \gg m_b$ for 
$\kappa \sim \kappa'$ one requires the condition
\begin{equation}
 \frac{Y_2}{Y_1} \approx -\frac{\kappa'}{\kappa}
\label{y12cond}
\end{equation}
to be satisfied. Since the lepton Yukawa couplings in 
equation~(\ref{y34cond}) are not governed by infrared quasi-fixed points 
and depend on the corresponding boundary conditions at the cutoff, the 
ratio of vacuum expectation values
\begin{equation}
\tan \beta \equiv \frac{\kappa}{\kappa'}
\end{equation}
is essentially a free parameter of the model. The full one-loop 
\RG equations for the Yukawa sector are found to be 
\bea
\SIXTPI\frac{dY_1}{dt} & = &5Y_1^3+7Y_1 Y_2^2 + Y_1(Y_3^2+Y_4^2)\nonumber\\
  & & \hspace{4em} + 2Y_2 Y_3 Y_4 
                 - \(8 g_3^2+\frac{9}{2}g_2^2+\frac{1}{6}g_1^2\)Y_1~,
\label{bet1}\\
\SIXTPI \frac{dY_2}{dt} & = &
5Y_2^3+7Y_2 Y_1^2 + Y_2(Y_3^2+Y_4^2) \nonumber \\
  & & \hspace{4em} + 2Y_1 Y_3 Y_4 
        - \(8 g_3^2+\frac{9}{2}g_2^2+\frac{1}{6}g_1^2\)Y_2~,
\label{bet2}\\
\SIXTPI \frac{dY_3}{dt} & = &
3Y_3^3 + Y_3 (Y_4^2 + Y_5^2) + 3 Y_3 (Y_1^2+Y_2^2) \nonumber \\
  & & \hspace{4em} + 6 Y_1 Y_2 Y_4 
        -\(\frac{9}{2}g_2^2+\frac{3}{2}g_1^2\)Y_3~,
\label{bet3}\\
\SIXTPI \frac{dY_4}{dt} & = &
3Y_4^3 + Y_4 (Y_3^2 + Y_5^2) + 3 Y_4 (Y_1^2+Y_2^2) \nonumber \\
  & & \hspace{4em} + 6 Y_1 Y_2 Y_3 
        -\(\frac{9}{2}g_2^2+\frac{3}{2}g_1^2\)Y_4~,
\label{bet4}\\
\SIXTPI \frac{dY_5}{dt} & = &
 \frac{7}{2}Y_5^3 + Y_5 (Y_3^2+Y_4^2)
              -\(\frac{9}{4}g_2^2+\frac{3}{4}g_1^2\)Y_5~.
\label{bet5}
\eea
For a heavy, non-propagating $\sigma$-scalar the coupling $Y_6$ decouples.
Evolving this set of Yukawa couplings down to low energies one finds
that now the square root $\sqrt{Y_1^2+Y_2^2}$ exhibits a fixed point behaviour
(just like $Y_1$ for the case $Y_2=0$), whereas the ratio of $Y_1$ and $Y_2$ 
runs very slowly (see figure~\ref{fig:y1y2plane}).
Thus the ratio $Y_2/Y_1$ at low energies depends in a straightforward way 
on the boundary condition at the cutoff. 
Unfortunately, the \RG evolution of Yukawa couplings does not 
automatically impose the relation~(\ref{y12cond}), since the boundary 
conditions on lepton and quark Yukawa couplings are unrelated when 
starting from the four-fermion Lagrangian. Thus, at this level one 
does not naturally explain the large top-bottom  mass splitting. 
However, since the ratios of Yukawa couplings run only slowly from high to low
energies, at least there exist sensible boundary conditions that lead,
with~(\ref{y34cond}), to the desired situation~(\ref{y12cond}) at low 
energies, which itself is a non-trivial point. 
Therefore one may assume that the relation~(\ref{y12cond}) is satisfied and 
leave it to the underlying physics to justify this choice.

Below the right-handed scale one should switch to the \SM 
$\beta$-functions of $Y_t$ and $Y_b$, which are obvious 
linear combinations of $Y_1$ and $Y_2$. Imposing the 
relation~(\ref{y12cond}) as argued above, one can find 
those values of $\tan \beta$ which lead to the correct top mass,
depending on the values of the cutoff $\Lambda$ and the right-handed
scale $\mu_R$. 
\begin{figure}[htb]
\centerline{
\epsfysize=78ex
\rotate[r]
{\epsffile{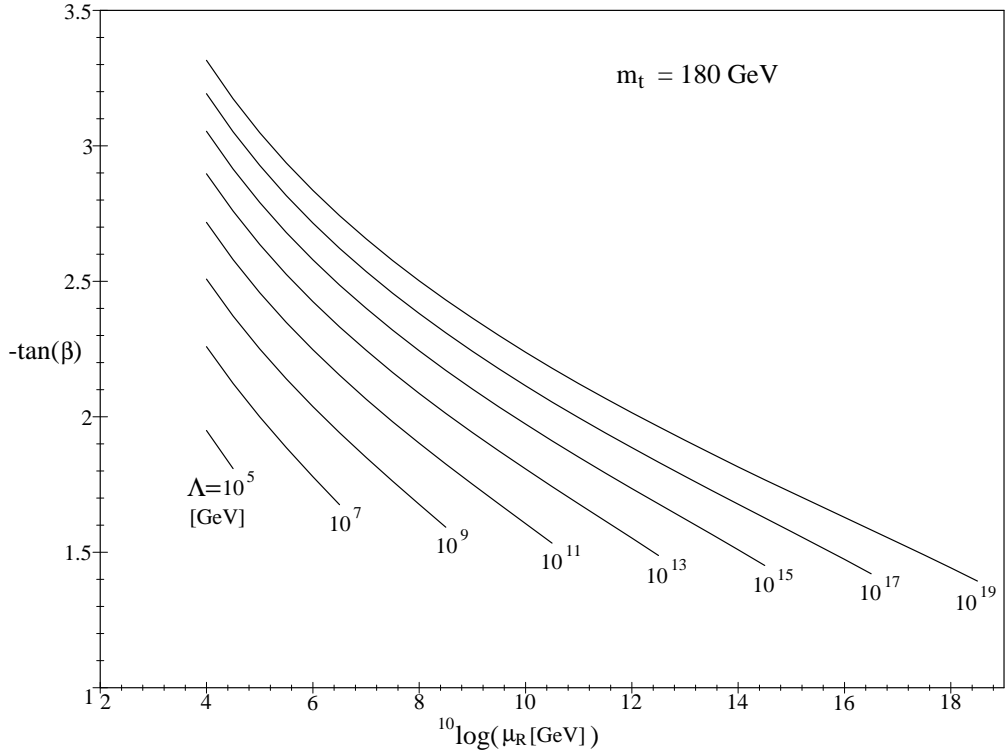}}}
\caption[]{\small\sl
Values of $\tan\beta$ for $m_t = 180$~GeV and various magnitudes of 
the cutoff $\Lambda$ and right-handed scale $\mu_R$.
\label{fig:figy12}}
\end{figure}

Figure~\ref{fig:figy12} shows the results for $\tan\beta$ 
assuming a top mass of $m_t=180$~GeV. One observes that a viable 
top mass can be obtained for a large range of possible values of 
the cutoff and for various parity breaking scales. This is in 
contrast with the top condensate approach to the \SM\cite{BHL}
where the lowest possible (but still too high) top mass arises for 
the largest possible cutoff. As one can see from the figure, this 
model can reproduce a viable top quark mass for values of the 
cutoff as low as $200$~TeV and for a parity breaking scale about 
$50$~TeV. This means that there is only a minimal amount of 
fine-tuning involved and the gauge hierarchy problem gets 
significantly ameliorated. If one considers different values
for the top mass, the whole set of curves in figure~\ref{fig:figy12} 
is slightly shifted vertically, e.g., for $m_t = (168 - 192)$~GeV
one finds an overall range for $\tan\beta$ of $(1.3 - 4.0)$.
Further phenomenological studies~\cite{ALSV1,ALSV2} showed that 
this model is compatible with all experimental data. 

Since the model was constructed in order to achieve a \SM decoupling 
limit, the phenomenological viability seems not too surprising. However, 
it is nontrivial to obtain a scenario where the symmetry breaking pattern 
works in the desired way and the top and Higgs mass predictions are
not in contradiction with the data. The model discussed above has 
altogether 9 input parameters (eight four-fermion couplings $G_1,\ldots G_8$ 
and the scale of new physics $\Lambda$), a factor of two less 
than the number of free parameters in conventional \LRS models.
It is thus constrained compared to the conventional model, and it 
was demonstrated that, e.g., for Higgs scalars in the triplet 
representation the correct pattern of symmetry breaking does not 
arise. We have shown that with scalar doublets the model allows the required 
features; in particular, parity breaking at low energies occurs
automatically regardless of the choice of the parameters of the model,
provided the gauge symmetry breaks at all. This is in striking contrast 
with conventional \LR models, in which for parity to be spontaneously 
broken the parameters of the Higgs potential must be chosen so as to 
satisfy a certain inequality. At the same time, the number of parameters 
to be tuned in order to achieve the correct {\em hierarchy} of the 
symmetry breaking scales is the same as in the conventional approach; 
in this respect the composite model has no advantages over conventional 
\LR models.

\section{Summary}

We emphasized that the \SM describes the existing data with remarkable 
accuracy. The success of the \SM implies that any theory of new 
physics will be phenomenologically viable if it has a suitable \SMLp 
By \SML we mean models which become, for certain parameter choices, 
indistinguishable from the \SMk and where the masses and couplings lie 
in the range allowed by the data. At the moment the \SML simply allows 
a scenario to hide behind the \SMk but it also ensures that potential 
deviations in the future can be understood as corrections to this limit.

Obviously the \SML implies that no extra light particles exist which
should have already been detected. There are further direct restrictions 
for extra couplings and masses which stem, for example, from rare decays 
and FCNC limits. Additional indirect limitations arise from radiative 
corrections. The agreement between the top mass value derived from 
radiative corrections with its experimental value leaves hardly any
room for further custodial SU(2) violating effects in the 
\EW precision variable $T$. This most likely requires that the \EW 
symmetry breaking operator is a doublet under $\mbox{SU(2)}_L$ with 
hypercharge $Y=1$. Due to non-decoupling effects this also disfavours 
models where extra fermionic doublets contribute to $T$ via loops. 
On the other hand, the ``size'' of the symmetry breaking sector
is limited by the \EW precision variable $S$, which roughly
counts the number of degrees of freedom. The smallness of $S$ probably 
points already towards the existence of a scalar Higgs particle as 
unitarity partner of the \GBsk since typically the contributions of 
vector like states (which mix and lead to 1/$M^2$ corrections) are 
considerably larger than those of scalars, which only
lead to logarithmic contributions. 
But even if there is a hint for a scalar Higgs particle, this 
does not yet tell us whether this scalar is composite or fundamental. 
Therefore, the current data contains no evidence about the nature of 
the solution to the hierarchy problem and the related new physics 
beyond the \SM at TeV scales. While the MSSM can be viewed as a theory 
with a \SMLk the difficulties with naive Technicolor are associated 
with the absence of such a limit. Some modern Technicolor approaches, 
such as Topcolor assisted Technicolor, essentially work in the 
direction of reestablishing such a \SMLp

Next we pointed out that it is possible to build models of \DSB 
which are systematically more viable due to a \SMLp We evaluated 
phenomenological guidelines which lead to such scenarios.
The most interesting aspect is that dynamics quite 
different from QCD is required in order to produce a scalar resonance 
instead of a rho-like vector. An example for a model which has a \SML 
is the so-called BHL model of \EW symmetry breaking. The BHL model is 
however unacceptable, since it cannot accommodate the correct top mass. 
With the help of the Pagels--Stokar relation we argued that this has 
systematic reasons and that the experimental top mass value is in 
general too small for a scenario with just a single top condensate. We 
postulated therefore a sequential breaking of an extended gauge group with 
a second condensate and more complex relations between the 
vacuum expectation values and masses. 

We presented a \LRS model which realizes the ideas
given above. This model is to our knowledge the first 
successful attempt to break \LR symmetry dynamically. We find a tumbling 
scenario where the breaking of parity and $\mbox{SU(2)}_R$ eventually 
drives the breaking of the \EW symmetry. The model gives a viable top 
quark mass value and exhibits a number of low and intermediate scale 
Higgs bosons. Furthermore it predicts relations between masses of various 
scalars and between fermion and Higgs boson masses which are in principle 
testable. If the right-handed scale $\mu_R$ is of the order of a few 
tens of TeV, the neutral CP-even and CP-odd scalars $\phi_{2r}^0$ 
and $\phi_{2i}^0$ can be even lighter than the \EW Higgs boson. In fact, 
they can be as light as $\sim 50-100$~GeV and thus might still be 
observable at LEP~II. To summarize, this model can be made consistent with 
all experimental data and demonstrates that viable models of \EW
symmetry breaking can be built. It should be interesting to investigate 
whether other models can be constructed along these lines. 

\section*{Acknowledgments}
We would like to thank D. Kominis, J. Manus and G. Triantaphyllou
for interesting and useful discussions. 
This work has been supported in part by DFG grant Li~519/2-2 and
in part by the Director, Office of Energy Research, Office of High 
Energy and Nuclear Physics, Division of High 
Energy Physics of the U.S. Department of Energy under Contract 
DE-AC03-76SF00098. E.S. is supported by a BASF research fellowship 
and the Studienstiftung des deutschen Volkes.



\begin{thebibliography}{99}
\bibitem{SMreview}  G. Altarelli, CERN-TH-96-265, Oct 1996, talk given at
                    Cracow International Symposium on Radiative Corrections 
                    (CRAD 96),
                    Cracow, Poland, Aug 1996, {hep--ph/9611239}. 
\bibitem{HP}        See e.g. M. Veltman, Acta Phys. Pol. B12 (1981) 437.
\bibitem{AWT}       See for example T. Appelquist, J. Terning and 
                    L.R.C.~Wijewardhan, 
                    Phys.~Rev.~Lett. 79 (1997) 2767.
\bibitem{SUGRA}     H. P. Nilles, Phys. Rep {110} (1984) 1.
\bibitem{gaugemed}  See e.g. 
                    M. Dine, A. E. Nelson, Y. Nir and Y. Shirman,
                    Phys. Rev. {D53} (1996) 2658;
                    H. Murayama, Phys. Rev. Lett. {79} (1997) 18. 
\bibitem{seib-intr} N. Seiberg, Phys. Rev. {D49} (1994) 6857; 
                    Nucl. Phys. {B435} (1995) 129; \\
                    For a review see e.g.  
                    K. Intriligator and N. Seiberg, 
                    Nucl. Phys. Proc. Suppl. 45BC (1996) 1.
\bibitem{haber97}   H.E. Haber, talk given at the 5th International Conference 
                    on Supersymmetries in Physics (SUSY 97), Philadelphia, 
                    May 1997, hep--ph/9709450.
\bibitem{mssmdecoupl} H.E. Haber, in {\it Perspectives for Electroweak 
                    Interactions in $e^+ e^-$ Collisions}, Proceedings of the 
                    Ringberg Workshop, Ringberg Castle, Tegernsee, 
                    5--8 February, 1995, edited by B.A. Kniehl 
                    (World Scientific, Singapore, 1995) pp. 219--231
\bibitem{mssmhiggs} H.E. Haber, R. Hempfling and A.H. Hoang, 
                    Z. Phys. {C75} (1997) 539;\\
                    M. Carena, J.R. Espinosa, M. Quiros and C.E.M. Wagner, 
                    Phys. Lett. {B355} (1995) 209;\\
                    M. Carena, M. Quiros and C.E.M. Wagner, 
                    Nucl. Phys. {B461} (1996) 407
                    and references therein. 
\bibitem{STUref}    M.E.~Peskin and T.~Takeuchi, Phys. Rev. Lett. 65 (1990) 
                    964, Phys. Rev. D 46 (1992) 381. 
\bibitem{TCruleout} See e.g. K. Lane, Proceedings of the International
                    Conference on High Eenergy Physics, Warsaw, Poland, 1996,
                    Eds. Z.~Ajduk and A.K.~Wroblewski, World Scientific, 1997;\\
                    R.S. Chivukula, talk presented at Advanced School on
                    Electroweak Theory, Mao, Menorca, June 1996, hep--ph/9701322;\\
                    K.~Lane, An Introduction to Technicolor, Proceedings of
                    the 1993 TASI School, Eds. S.~Raby and T.~Walker,
                    World Scientific, 1994.
\bibitem{Weinberg}  S. Weinberg, Phys. Rev. Lett. 18 (1967) 507.
\bibitem{largeNc}   G. 't Hooft, Nucl. Phys. B72 (1974) 461.
\bibitem{Sexp}      P.~Langacker and J.~Erler, hep--ph/9703428 and 
                    {\em Review of Particle Properties}, Phys. Rev. D 54 
                    (1996) 1.
\bibitem{bestH}     See e.g. W. Hollik, Heavy Flavours II, ed. by A.J. Buras
                    and M. Lindner, World Scientific, Singapore, 1998. 
\bibitem{BlLi}      A. Blumhofer and M. Lindner, Nucl. Phys. B407 (1993) 173.
\bibitem{ETC}       S. Dimopoulos and L. Susskind, Nucl. Phys. B155 (1979) 
                    237;\\
                    E. Eichten and K. Lane, Phys. Lett. B90 (1980) 125. 
\bibitem{AppWi}     T. Appelquist, T. Takeuchi. M.B. Einhorn and
                    L.C.R.~Wijewardhana, Phys. Lett. B232 (1989) 211.
\bibitem{BHL}       W.A.~Bardeen, C.T.~Hill and M. Lindner, Phys. Rev. D41 
                    (1990) 1647.
\bibitem{N}         Y. Nambu, in {\em New Theories in Physics, Proc. XI Int.
                    Symposium on Elementary Particle Physics}, eds. Z. Ajduk,
                    S. Pokorski and A. Trautman (World Scientific, Singapore,
                    1989) and EFI report No. 89-08 (1989), unpublished.
\bibitem{Mir}       A. Miransky, M. Tanabashi and K. Yamawaki,
                    Mod. Phys. Lett. A4 (1989) 1043; Phys. Lett.  B221 (1989)
                    177.
\bibitem{Mar}       W.J. Marciano, Phys. Rev. Lett. 62 (1989) 2793.
\bibitem{VL}        V.G. Vaks and A.I. Larkin, Sov. Phys. JETP 13 (1961) 192.
\bibitem{NJL}       Y. Nambu and G. Jona-Lasinio, Phys. Rev. 122 (1961) 345.
\bibitem{CJT}       J.M. Cornwall, R. Jackiw and E. Tomboulis, Phys. Rev. D10 
                    (1974) 2445.
\bibitem{Bl}        A. Blumhofer, Phys.Lett. B320 (1994) 352; Nucl.Phys.
                    B437 (1995) 25.
\bibitem{IRqFP}     C.T. Hill, Phys. Rev. D24 (1981) 691;\\
                    C.T. Hill, C.N. Leung and S. Rao, Nucl. Phys. B262 (1985) 
                    517;\\
                    J. Bagger, S. Dimopoulos and M. Masso, Phys. Lett. 156B 
                    (1985) 357.
\bibitem{topcolor}  C.T. Hill, Phys. Lett. B266 (1991) 419;\\
                    For a review see C.T. Hill, invited talk at International 
                    Workshop on Perspectives of Strong Coupling
                    Gauge Theories (SCGT 96), Nagoya, Japan, 13-16 Nov
                    1996, {hep--ph/9702320}.
\bibitem{topcolor2} S.P. Martin, Phys. Rev. D45 (1992) 4283.  
\bibitem{U1}        M. Lindner and D.~Ross, Nucl. Phys. B370 (1992) 30.
\bibitem{SU2V}      R. B\"onisch, Phys. Lett. B268 (1991) 394.
\bibitem{PS}        H. Pagels and S. Stokar, Phys. Rev. D20 (1979) 2947;\\
                    A. Carter and H. Pagels, Phys. Rev. Lett. 43 (1979) 1845;\\
                    R. Jackiw and K. Johnson, Phys. Rev. D8 (1973) 2386.
\bibitem{LR1}       J.C.~Pati and A.~Salam, Phys. Rev. D10 (1975) 275;\\
                    R.N.~Mohapatra and J.C.~Pati, Phys. Rev. D11 (1975) 566;
                    2558; \\
                    G.~Senjanovi\'c and R.N.~Mohapatra, Phys. Rev. D12 (1975) 1502.
\bibitem{LR3}       R.N. Mohapatra and G. Senjanovi\'c, Phys. Rev. Lett. 44
                    (1980) 912; Phys. Rev. D23 (1981) 165.
\bibitem{ALSV1}     E.~Akhmedov, M. Lindner, E.~Schnapka and J.~Valle, 
                    Phys. Lett. B368 (1996) 270.
\bibitem{ALSV2}     E.~Akhmedov, M. Lindner, E.~Schnapka and J.~Valle, 
                    Phys. Rev. D53 (1996) 2752.
\bibitem{Auxform}   T.~Eguchi, Phys.~Rev.~D14 (1976) 2755;\\
                    F.~Cooper, G.~Guralnik and N.~Snyderman,
                    Phys.~Rev.~Lett.~40 (1978) 1620.
\bibitem{LR2}       G. Senjanovi\'c and R.N. Mohapatra, ref.~\cite{LR1}
\bibitem{WW}        D. Wyler and L. Wolfenstein, Nucl. Phys. B218 (1983) 205.
\bibitem{LR}        See e.g. G.~Senjanovi\'{c}, Nucl. Phys. B153 (1979) 334;\\
                    C.S.~Lim and T.~Inami, Prog. Theor. Phys. 67 (1982) 1569;\\
                    F.I.~Olness and M.E.~Ebel, Phys. Rev. D32 (1985) 1769;\\
                    J.F.~Gunion, J.~Grifols, A.~Mendez, B.~Kayser and F.~Olness,
                    Phys. Rev. D40 (1989) 1546.
\bibitem{CDFD0}     P.L. Tipton, Proceedings of the 28th International 
                    Conference of High Energy Physics, Warsaw 1996, 
                    Eds. Z. Ajduk and A.K. Wroblewski, World Scientific.
\bibitem{LRdesh}    N.G. Deshpande, J.F. Gunion, B. Kayser and F. Olness,
                    Phys. Rev. D44 (1991) 837.
\bibitem{quasiIR}   C.T.~Hill, Phys. Rev. D24 (1981) 691.
\bibitem{manus}     A. Blumhofer, R. Dawid and J. Manus, hep--ph/9710495.
\bibitem{latticeFP} See e.g.
                    B.~Lin, I.~Montvay, G.~M\"unster, M.~Plagge and 
                    H.~Wittig, Phys. Lett. B317 (1993) 143.
\bibitem{oneloopb}  See e.g. T.P.~Cheng, E.~Eichten and L.F.~Li, 
                    Phys. Rev. D9 (1974) 2259.
\end{thebibliography}
\end{document}